\documentclass{aa}

\usepackage[version=3]{mhchem}
\usepackage{graphicx}
\usepackage{subfigure}
\usepackage{amsmath}

\usepackage{txfonts}
\usepackage{natbib}
\bibpunct{(}{)}{;}{a}{}{,} 

\newcommand{\bi}{\begin{itemize}}
\newcommand{\ei}{\end{itemize}}
\newcommand{\be}{\begin{enumerate}}
\newcommand{\ee}{\end{enumerate}}

\newcommand{\beq}{\begin{equation}}
\newcommand{\eeq}{\end{equation}}

\newcommand{\five}{\ce{H2O}, CO, \ce{CO2}, \ce{CH4}, and \ce{O2}}

\begin{document}

\title{Molecular abundances and C/O ratios in chemically evolving planet-forming disk midplanes}

\author{Christian Eistrup \inst{1}
\and Catherine Walsh\inst{1,2}
\and Ewine F. van Dishoeck\inst{1,3}}

\institute{Leiden Observatory, Leiden University, P.O. Box 9513, 2300 RA Leiden, the Netherlands\\
\email{eistrup@strw.leidenuniv.nl}
\and School of Physics and Astronomy, University of Leeds, Leeds LS2 9JT, UK\\
\email{c.walsh1@leeds.ac.uk}
\and Max-Planck-Institut f\"ur Extraterrestrische Physik, Giessenbachstrasse 1, 85748 Garching, Germany\\
\email{ewine@strw.leidenuniv.nl}}

\date{Received $\cdots$ / Accepted $\cdots$}

 
\abstract
{Exoplanet atmospheres are thought be built up from accretion of gas as well as pebbles and planetesimals in the midplanes of planet-forming disks. The chemical composition of this material is usually assumed to be unchanged during the disk lifetime. However, chemistry can alter the relative abundances of molecules in this planet-building material.}
{To assess the impact of disk chemistry during the era of planet formation. This is done by investigating the chemical changes to volatile gases and ices in a protoplanetary disk midplane out to 30 AU for up to 7 Myr, considering a variety of different conditions, including a physical midplane structure that is evolving in time, and also considering two disks with different masses.}
{An extensive kinetic chemistry gas-grain reaction network is utilised to evolve the abundances of chemical species over time. Two disk midplane ionisation levels (low and high) are explored, as well as two different makeups of the initial abundances (``inheritance'' or ``reset'').}
{Given a high level of ionisation, chemical evolution in protoplanetary disk midplanes becomes significant after a few times $10^{5}$ yrs, and is still ongoing by 7 Myr between the H$_{2}$O and the O$_{2}$ icelines. Inside the H$_{2}$O iceline, and in the outer, colder regions of the disk midplane outside the O$_{2}$ iceline, the relative abundances of the species reach (close to) steady state by 7 Myr. Importantly, the changes in the abundances of the major elemental carbon and oxygen-bearing molecules imply that the traditional ``stepfunction'' for the C/O ratios in gas and ice in the disk midplane (as defined by sharp changes at icelines of H$_{2}$O, CO$_{2}$ and CO) evolves over time, and cannot be assumed fixed, with the C/O ratio in the gas even becoming smaller than the C/O ratio in the ice. In addition, at lower temperatures (< 29 K), gaseous CO colliding with the grains gets converted into CO$_{2}$ and other more complex ices, lowering the CO gas abundance between the O$_{2}$ and CO thermal icelines. This effect can mimic a CO iceline at a higher temperature than suggested by its binding energy.}
{Chemistry in the disk midplane is ionisation-driven, and evolves over time. This affects which molecules go into forming planets and their atmospheres. In order to reliably predict the atmospheric compositions of forming planets, as well as to relate observed atmospheric C/O ratios of exoplanets to where and how the atmospheres have formed in a disk midplane, chemical evolution needs to be considered and implemented into planet formation models.}

\keywords{protoplanetary disks -- planet formation -- astrochemistry -- planetary atmospheres -- molecular processes}

\titlerunning{Molecular abundances and C/O ratios in chemically evolving...}
\maketitle


\section{Introduction}
\label{intr}

Over the past 20 years exoplanetary science has evolved from purely detection 
\citep[e.g.,][]{udry2007,borucki2011,batalha2013,fischer2014} to 
exoplanet characterisation. 
One example of characterisation is the deciphering of the molecular content of 
exoplanet atmospheres which has revealed simple molecules 
such as CO and \ce{H2O}, and possibly \ce{CO2} and \ce{CH4}  
\citep[e.g.,][]{seager2010,snellen2010,birkby2013,fraine2014,crossfield2015,sing2016}.

The C/O ratio of an exoplanet atmosphere (usually derived from
relative abundances of the simple molecules listed above) has been
proposed to be a possible tool to link an exoplanet to its formation
site in the natal protoplanetary disk.  This is because disk midplanes
have a radial gradient of C/O ratios in gas and ice 
\citep[see e.g.][]{oberg2011co}.
This approach, to couple observed C/O ratios in exoplanets with their 
possible formation sites, typically assumes that the radial dependence
of the midplane C/O ratio is defined solely by the iceline (snowline)
positions of the main volatile species in a disk of fixed chemical
composition \citep[see e.g.][]{marboeuf14}. The position of the iceline for a particular species depends on the
temperature structure of the disk midplane, and the binding energy (or
desorption energy) of the molecule to ice mantles on dust grains.
Icelines thus affect the relative ratios of heavy elements in the gas
and ice. For example, beyond the \ce{H2O} iceline, the gas is depleted in
oxygen whilst the ice is enriched in oxygen.  Similar changes occur
across the \ce{CO2}, \ce{CH4}, and CO icelines thereby also altering the local
C/O ratio in both the ice and gas.  An example of a variable C/O ratio
for a disk with a fixed chemical composition (as per the conventional assumption) is shown in Fig.~1 of
\citet{oberg2011co}. Due to the step-like features of that figure, caused by icelines, the figure will be referred to as the traditional ``stepfunction''.


Protoplanetary disks are not static objects: over the disk lifetime, 
from formation to dispersal, typically $\sim 0.1 - 10$~Myr, 
the disk generally cools, spreads, loses material (via accretion onto the star or 
dissipation) and thus becomes less dense over time \citep[see the review by][]{williams11}.  
Hence, the icelines of volatile species also change in time, 
generally migrating inwards towards the star \citep{davis2005,min2011,harsono2015}. To date, most models exploring the location and evolution of icelines
with the explicit goal to link to the composition of planet-building
material have assumed static midplane conditions \citep{oberg2011co,madhu2014,thiabaud2015gascomp}.
Static models are more straightforward to couple with chemical
kinetics.  Recently \citet{piso2015,piso2016} explored the evolution
of elemental ratios across icelines in both a static and viscously evolving
disk model including the radial drift of ice-coated grains of
different sizes.  The temperature structure is kept fixed in time in
these models. A similar approach has been adopted by \citet{alidib2014} and \citet{alidib2017}. Both of these models include dust and pebble
dynamics at the expense of kinetic chemistry.  Those models which do
allow evolving disk midplane conditions also neglect kinetic chemistry
for simplicity \citep[see, e.g.,][]{marboeuf14,mordasini2016}.


The impact of kinetic chemistry in planet formation models is yet to be fully explored due to the increased computational demands in coupling a chemical model with e.g., a planet population synthesis model. This is despite chemical kinetics
being a necessity in models explaining protoplanetary disk
observations for some 20 years since the first detection of molecules,
other than CO, in disk atmospheres \citep[see review by][]{henning13}. This is now
changing, as is the traditional picture of a static disk. Recent examples of kinetic gas-grain chemistry in disks can be found in \citet{helling2014}, \citet{walsh15} and \citet{cridland2016}, with different levels of treatment of the chemistry and grain growth. Also, there is growing observational evidence for low gaseous CO abundances which can be attributed to either rapid planetesimal growth or chemical evolution \citep[see e.g.][]{favre2013,bruderer2012,reboussin2015,kama2016,miotello2017}. 

The disk midplanes suffer from a lack of observational
constraints on their chemical compositions.  This due to the
observational difficulty of probing the midplane through the optically
thick higher-lying layers of the disk and to the freeze-out of species
in the outer disk midplane. Most molecular emission, whether at
millimeter or infrared wavelengths, therefore arises from the warm
layer at intermediate disk heights \citep[see][for a
review]{bergin2007}. Even the \ce{N2H+} emission observed with ALMA, which probes
the part of the outer disk where CO is frozen out, arises mostly from
layers just above the midplane \citep{qi2013,hoff2017}. Molecules expected to be abundant in midplane ices such as \ce{H2O} \citep{hogerheijde2011,du2017}, \ce{NH3} \citep{salinas2016}, \ce{H2CO} \citep{loomis2015} and \ce{CH3OH} \citep{walsh2016}
have been detected in the gas-phase but the observations either lack
spatial resolution or reflect subsequent gas-phase chemistry, making
it difficult to infer the outer disk midplane ice abundances. Ultimately,
comet abundances may provide the best constraints on this region, at
least for our solar nebula disk.

In contrast, the midplane of the warm inner disk where CO is not
frozen out has recently been imaged for the first time through CO
isotopologue observations of \ce{^{13}C^{18}O} by \citet{zhang2017} for one
disk. Such observations are promising, and are perhaps also feasible
for other common species, but they require large telescope time
investments. Also, the dust continuum emission becomes optically thick
in the inner disk, hiding our view of much of the planet-forming
midplane. Thus, perhaps ultimately the only information on the
midplane chemical composition is to be inferred from the atmospheric
compositions of the planets that form out of the midplane, and which
remain after the disk is dispersed.

As emphasized in \citet{mordasini2016}, the road from a
gaseous disk midplane to an observable exoplanet atmospheric
composition is long, requiring many steps, physical (e.g., migration)
as well as chemical.  A key issue is whether the planets' heavy element
enrichment is controlled by the accreted gas or by the accreted (icy)
planetesimals. \citet{mousis2009}, \citet{thiabaud2015elemrat}, and \citet{mordasini2016} found that
the dominant delivery mechanism for heavy elements to a forming
gas-giant planet atmosphere is accretion of icy planetesimals rather
than direct accretion from the gas, at least for planets with masses
less than a few times that of Jupiter. The
ultimate C/O ratio in the observable part of the planet's atmosphere
also depends on how well the sublimated planetesimals are mixed in the
envelope and on the core-envelope partitioning. In any case, it is
clear from models such as those of \citet{mordasini2016}, which
explicitly include planetary formation and planetesimal accretion
processes, that the composition of both the gas and the ice of the
evolving parent disk is needed to determine the planetary atmosphere
composition.





\citet{eistrup2016} (herefrom Paper 1) investigated how kinetic chemical
evolution affects the composition of volatiles (ice and gas) in a
static protoplanetary disk midplane. If
interstellar ices are inherited from the collapsing cloud (without
alteration) by the protoplanetary disk, then the ice composition is
preserved only for the case where there is no source of external
ionisation (i.e., no cosmic rays). An important conclusion from Paper 1 is that if chemistry is efficient on a timescale of $\sim 1$~Myr,
then the C/O ratio of the gas and ice remains $< 1$ always within the
CO iceline.  Only for the case of low ionisation and full inheritance
does the C/O ratio approach 1, and this only occurs for the gas: the
ice remains oxygen-rich always.  



The effects of chemical kinetics on the composition of gas \emph{and}
ice volatiles in evolving protoplanetary disk midplanes remains to be
quantified.  We here expand upon the work presented in Paper 1 to investigate
how changing temperature and density in the disk midplane with time
modifies the above conclusions. We also consider how disk mass may
affect the composition of planet-building material.  We furthermore
study the timescale for chemical changes by extending the lifetime of
our disk models to $\sim 7$~Myr at which point we assume that
planet formation has halted and disk dispersal is advanced.

We investigate a wide parameter space, with somewhat simplified physical structures, and two sets of extreme initial abundances. The two scenarios of initial chemical abundances are: cloud inheritance (hereafter ``Inheritance'' scenario) and chemical reset (hereafter ``Reset'' scenario). In the inheritance scenario the initial abundances are molecular, and taken to be those found in ices in the parent molecular cloud. In the reset scenario, the initial abundances are atomic, because all molecules are assumed dissociated by high temperatures close to the young star. Both scenarios include H, \ce{H2} and He initially, with abundance ratios: $N$(H)/$N$(\ce{H2})=10$^{-4}$, and $N$(He)/$N$(\ce{H2})=0.17. Generally, the inner disk midplane should be susceptible to luminosity outbursts that can reset the chemistry, while the outer disk midplane should be expected to inherit the parent cloud abundances, so both scenarios may apply but in different parts of the disk \citep{pontoppidan2014}. The intention is to investigate chemical trends expected for these two sets of initial conditions and using a more realistic disk, which is becoming cooler and more diffuse over time.

\section{Methods}
\label{methods}

The evolving physical disk model from \citet{alibert13} is used. This disk model has been adopted in planet population synthesis studies. It features time evolution of disk midplane temperature and
pressure as a function of radius ($0.2 - 30$~au) over a timescale of 7
Myr.  
The approach for calculating the ionisation rate and chemical evolution 
follows that presented in Paper 1: we summarise the main 
aspects here. 



\subsection{Evolving disk model}
\label{evolving_adapt}

An evolving disk here refers to a disk that is cooling in time whilst
also losing mass. In this work power laws are fitted to the temperature 
structure evolving in time as originally computed by \citet{alibert13}. 

\begin{figure*}
\subfigure{\includegraphics[width=0.5\textwidth]{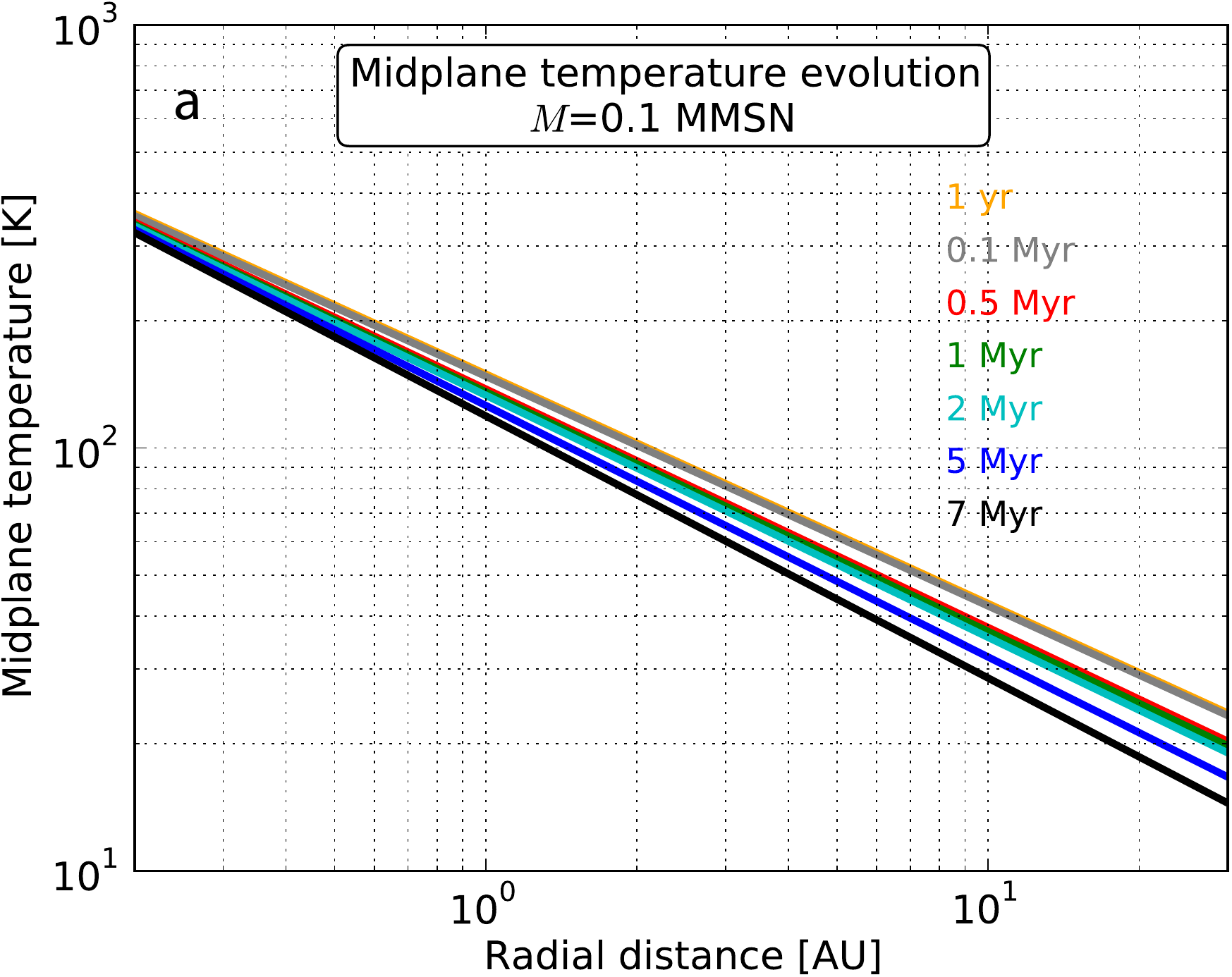}}
\subfigure{\includegraphics[width=0.5\textwidth]{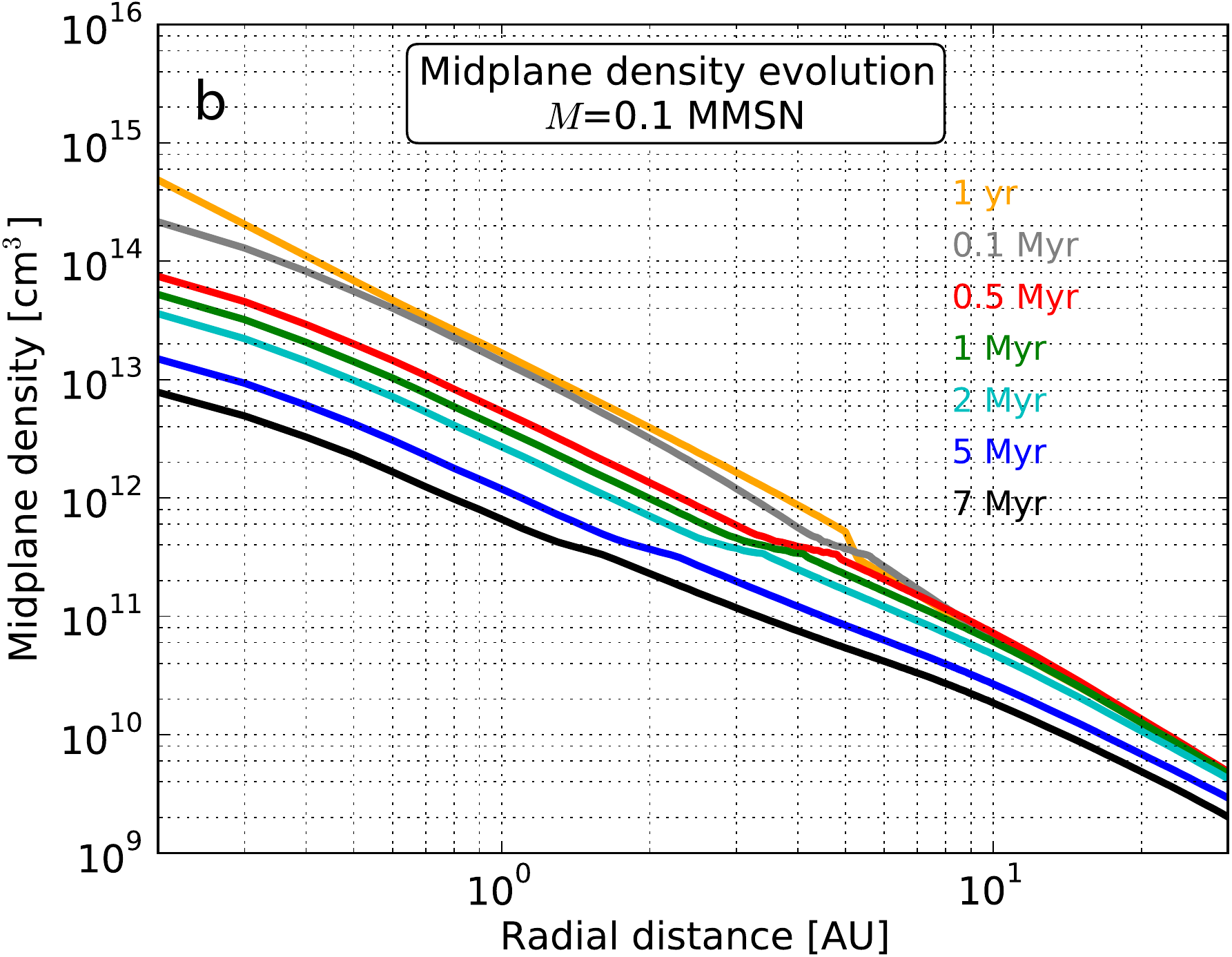}}\\
\subfigure{\includegraphics[width=0.5\textwidth]{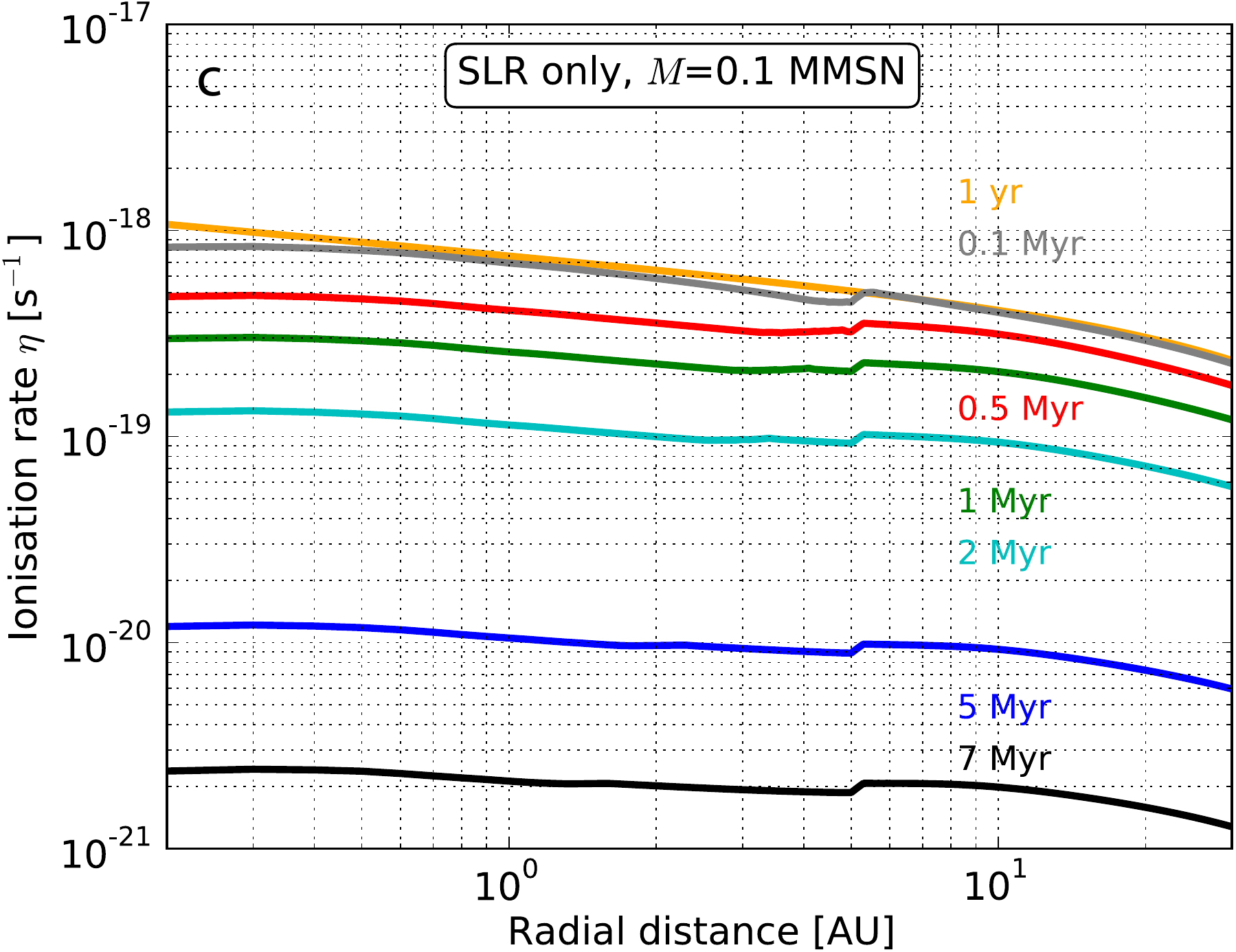}}
\subfigure{\includegraphics[width=0.5\textwidth]{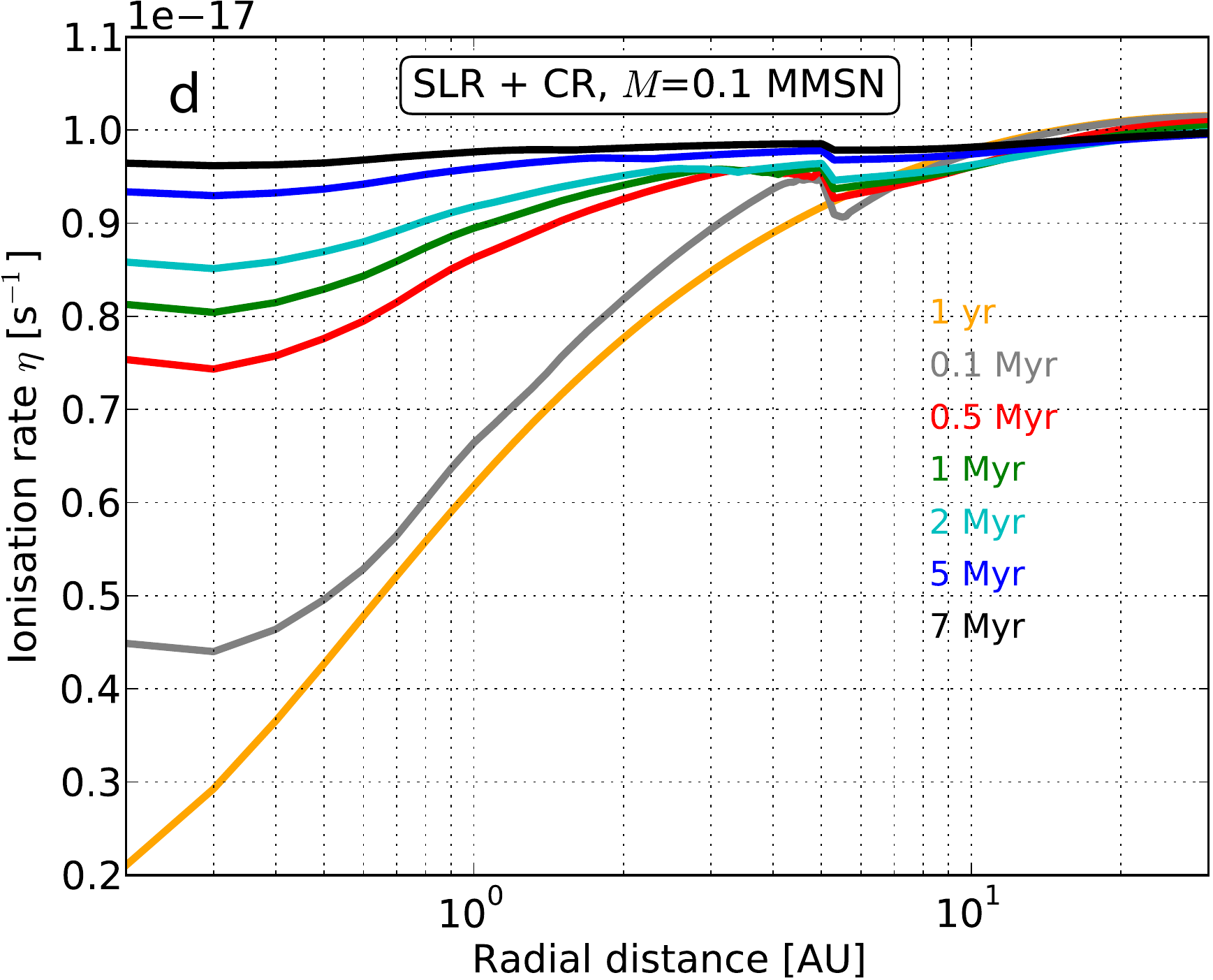}}\\
\caption{Midplane temperature and density (top) and ionisation rates 
(bottom) as functions of disk radius for different evolutionary times 
for the 0.1 MMSN evolving disk model. For this model, the starting conditions 
at 1 yr are given by the orange profiles shown here, which feature slightly 
different physical conditions than that for the static disk structure 
adopted by Paper 1 (see text for details).}
\label{low_mass_phys}
\end{figure*}

The disk structure adopted from \citet{alibert13} and used in Paper 1 will be
referred to as the 0.1 MMSN disk, having an initial mass of
$1.3\times 10^{-3}M_{\odot}$ or 0.13 times the minimum mass solar nebula, or MMSN \citep{weidenschilling1977}. The reason for fitting power laws is that
the initial time-dependent temperature structure from \citet{alibert13} featured a step, i.e., an unphysical radial drop, at around 5
AU, as well as an artificial lower temperature cut-off at 20 K. Both of
these features were smoothed out with a power law profile in Paper 1.
The same unphysical drop in temperature is observed for later time
steps. In order to smooth out all timesteps, whilst ensuring a
temperature profile that is monotonically decreasing in time for all
radii, power laws are fitted to all temperature profiles from
\citet{alibert13}.  For timesteps up to 400~kyr of evolution, the
power laws are fitted using model data from 1 to 2.1 AU, and for all
later timesteps from 0.2 to 0.6 AU.  The resulting power law fits are
then extrapolated to the entire radial range of the disk.  This
procedure gives the temperature profiles that can be seen for selected
timesteps (as labeled) in Fig.~\ref{low_mass_phys}a.

Using these temperature profiles, the midplane density profiles for
all timesteps are computed using the midplane pressure profiles from
\citet{alibert13}, assuming \ce{H2} to be the main gas
constituent. The resulting midplane number density radial profiles at
selected timesteps are shown in Fig.~\ref{low_mass_phys}b.

The surface density profiles at the first timestep are computed using
the prescription and values for Table 1, Disk ``8'' in \citet{alibert13}. Taking into account
disk evolution, the surface density at later timesteps at a radial
distance $R$ is computed by scaling the initial surface density
$\Sigma_{t=0}(R)$ to the evolution in midplane number density $n(R)$
as seen in Fig.~\ref{low_mass_phys}b. 
Hence, the surface density
at point $R$ at time $t$ is given by
\beq
\Sigma(R,t) = \Sigma(R,t=0) \times \frac{n(R,t)}{n(R,t=0)}
\label{sigma_evol}
\eeq

To drive chemical evolution, ionisation of the disk midplane is
needed, a conclusion reached by our and other models \citep[see e.g.][]{helling2014,aikawa96,willacy1998,walsh2012,cleeves13slr}. 
Paper 1 found that chemical evolution was observed to be significant within
1 Myr evolution time only when short-lived radionuclides
(SLRs) \emph{and} cosmic rays (CRs) were included.  
In this paper, the contribution to ionization by SLRs and CRs are 
treated in a similar way to that in Paper 1: 
the ionisation rate from decay of SLRs is computed using the 
prescription from Eq.~30 in \citet{cleeves14}, 
including the time-dependent term, which was omitted in Paper 1.


\beq
\zeta_{\ce{SLR}}(R,t)=(2.5 \cdot 10^{-19}\mathrm{s}^{-1}) 
\left(\frac{1}{2}\right)^{1.04t}\left(\frac{\Sigma(R,t)}{\mathrm{g~cm}^{-2}}\right)^{0.27}
\label{eq_slr}
\eeq

The contribution from CRs is calculated assuming attenuation
depending on how much material the CRs have to pass through on their
way to the midplane. 
The attenuation factor is calculated using the
(now) evolving surface density of the disk in the vertical direction 
giving the following ionisation rate

\beq
\zeta_{\ce{CR}}(R,t) = \zeta_{0}\times \mathrm{exp} \left(\frac{\Sigma(R,t)}{96 \mathrm{g~cm}^{-2}}\right),
\eeq

where $\Sigma(R,t)$ is the evolving surface density profile, taken from Eq. \ref{sigma_evol}, and $\zeta_{0}$=10$^{-17}$ s$^{-1}$ (as in Paper 1).

The SLR ionisation rate profiles (``low'' ionisation) and the
SLR + CR ionisation contribution profiles (``high'' ionisation) are
plotted at selected timesteps in Fig.~\ref{low_mass_phys}c and
d. Note that the SLR ionisation level decreases rapidly with time and
becomes very low, of order 10$^{-22}$~s$^{-1}$ by $\sim 7$~Myr. 
In contrast, the CR ionisation level increases with time, 
especially in the inner disk, due to the decreasing surface density.

The timestep grid of \citet{alibert13} is adopted, with physical conditions 
changing for every timestep of their models. 
However, timesteps are added at early evolutionary stages ($1-10^{4}$ yrs) 
in order to account for short-term chemical effects. 
The first timestep in the time grid in this work is 1 yr, and the physical 
conditions at 1 yr from \citet{alibert13} are adopted for this timestep. 
Four logarithmic decades are applied ranging from 1 to 10$^{4}$ yrs, with 
physical conditions during this time fixed at the conditions at 1 yr. 
After 10$^{4}$ yrs, the time grid and the associated, continuously changing 
physical conditions of \citet{alibert13} are adopted. 
The time goes up to 7 Myr, in order to also investigate long-term 
chemical evolution. 
Until 7 Myr, this disk model goes through 401 timesteps.

\subsection{Higher disk mass: 0.55 MMSN disk}

For investigating the effect of a higher disk mass, a 
more massive evolving disk physical structure from \citet{alibert13} 
is adopted \citep[Table 1, ``Disk 12'' of][]{alibert13}. The initial mass of this disk 
is 0.55 MMSN = 0.0055 $M_{\odot}$, determined in the same manner as was the case for the less massive disk in Paper 1.
Since this disk has the same steep temperature drop as was the case 
with the 0.1 MMSN disk, and also features the 20~K lower temperature 
limit, power laws are also fitted to the temperature profiles at each 
timestep. In order to have the temperature profiles decreasing in time 
monotonically at each radial point, for this disk, the power law fitting 
is done for the first timestep (1 yr) and the last timestep (7 Myr) 
for the evolving disk temperature profiles between 1.1 and 1.5 AU. 
For all intermediate timesteps, power law indices intermediate to the 
two fitted indices are adopted, so that the power law index is continuously 
and linearly decreasing in time, with the step-wise decrease in index 
value being constant for each timestep. The original disk temperature 
structure from \citet{alibert13} is then adopted inside 1.1 AU,  
whereas the fitted (for first and last timesteps) and imposed power law 
structures are applied outside 1.1~AU.
For the outer parts of the disk, at late timesteps, the temperature gets 
very low (below 8~K), which is unphysical. 
A lower limit of 8~K to the temperature is therefore imposed.


The midplane number density and ionization level 
structures for this 0.55 MMSN disk are computed in the same way as described for the 0.1 MMSN disk 
in Sect.~\ref{evolving_adapt}. The resulting midplane temperature, density 
and ionisation profiles for the midplane of the 0.55 MMSN disk are presented in 
Fig.~\ref{high_mass_phys} (see Appendix). The evolving 0.55 MMSN disk goes through 438 time 
steps during the 7~Myr modeled evolution. 
The first timestep is at 1 yr, and from 1 yr through 10$^{4}$ yrs the time resolution is the same as for the same early timesteps for the 0.1 MMSN disk model, as described in Section \ref{evolving_adapt}. After 10$^{4}$ yrs, the timesteps from \citet{alibert13} are adopted.

The disk structures differ for the two different disk masses. Since the ionisation level is dominated by the contribution from CRs at all times throughout the disk (see Fig. \ref{low_mass_phys} and \ref{high_mass_phys}, panels c and d), the decreasing surface density means the degree of ionisation in the inner disk is increasing in time. Because of the higher densities, the degree of ionisation in the inner disk is increasing faster in time for the 0.55 MMSN disk than for the 0.1 MMSN disk.

The temperature structure for the 0.55 MMSN disk is hotter in the inner disk, and has a steeper radial gradient, than that for the 0.1 MMSN disk. The low temperature in the outer disk for the 0.55 MMSN disk is due to lesser heating from irradiation from the central star. This is because more material shields the radiation for the 0.55 MMSN disk than for the 0.1 MMSN disk.

\begin{figure}
\subfigure{\includegraphics[width=0.5\textwidth]{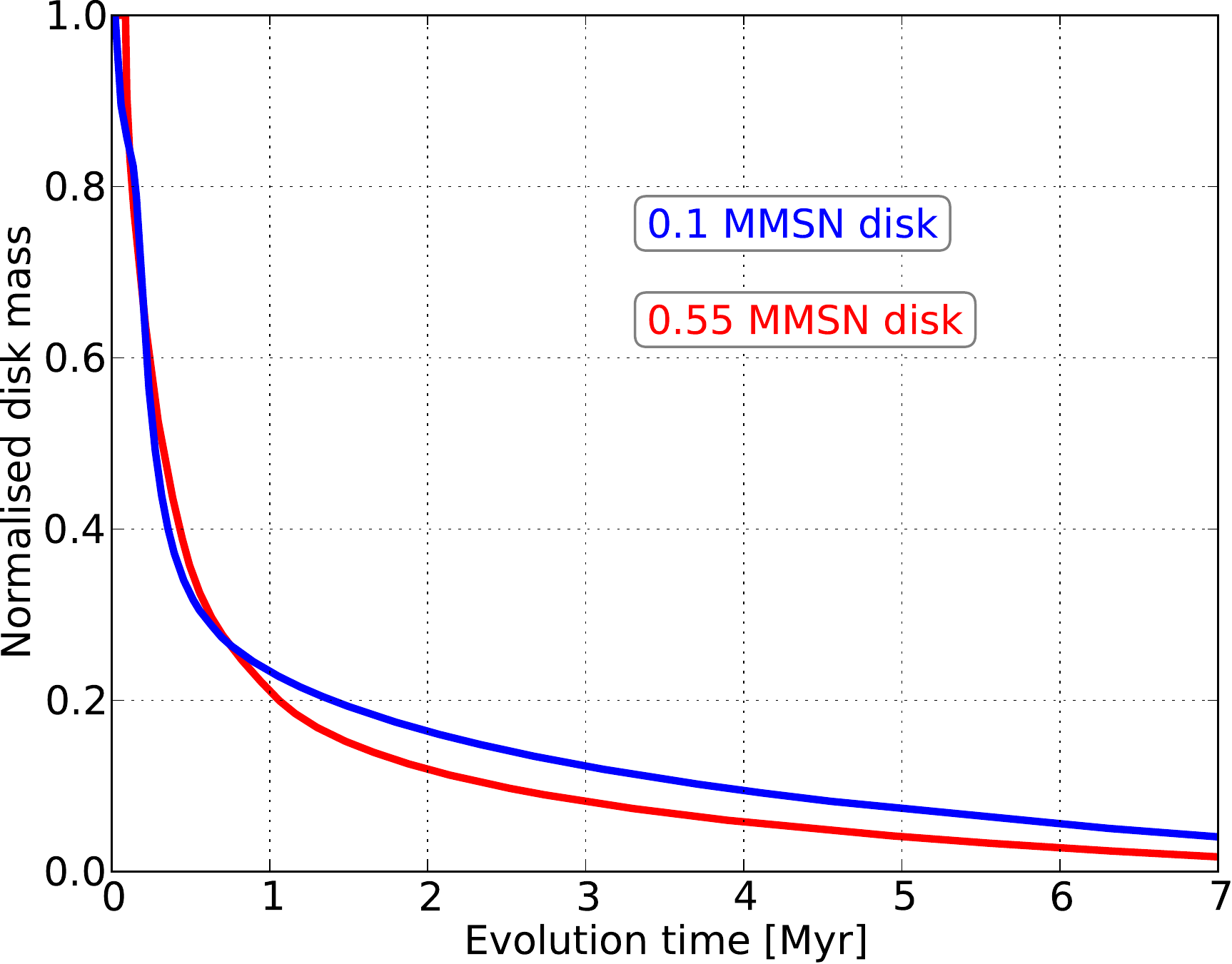}}
\caption{Evolving disk masses for the two disks, normalised to the initial disk masses of 0.1 and 0.55 MMSN. For both disks, more than half of the initial mass is lost before 0.5 Myr, and by 1-1.5 Myr evolution, only 20\% of the initial masses remain.}
\label{evol_diskmass}
\end{figure}

Fig. \ref{evol_diskmass} shows the evolution of the disk masses for the two disks, normalised to their initial masses. Both disks experience significant mass loss during the evolution, with the 0.1 MMSN disk losing ~95\% of its initial mass, and the 0.55 MMSN disk losing ~98\% of its initial mass, after 7 Myr.

\subsection{Chemical model}
\label{chem_model}
The chemical model includes gas-phase chemistry, gas-grain interactions
and grain-surface chemistry.  The gas-phase chemistry is from the
latest release of the UMIST Database for Astrochemistry
\citep[][]{mcelroy13} termed {\sc Rate}12.  
The rates for gas-grain interactions and grain-surface chemistry are calculated 
as described in \citet[][and references therein]{walsh15}.  
A gas/dust mass ratio of 100 is adopted. 
The chemical model used here is identical to the ``Full chemistry'' setup in
Paper 1. Photodissociation and photodesorption due to
cosmic-ray induced UV photons are included.

The adopted grain size is 0.1~$\mu$m. As discussed in Paper 1, this
assumption is certainly not correct, but the effects of grain growth
to larger sizes and decreasing midplane gas/dust ratios due to
settling tend to compensate each other. The important dust parameter for the chemistry is the amount of surface area available for freezeout and subsequent grain-surface chemistry.  This is often parameterised as a combination of grain size and number density (set by the assumed gas-to-dust mass ratio). The assumption on the constant grain size will be
further discussed in Section \ref{caveat_grains}. More generally, all trends found in this
paper are robust but the exact timescale for chemical changes and the
magnitude of the chemical changes will depend on the assumed grain size and
the gas/dust mass ratio. 

For details about the two sets of initial abundances (the ``inheritance'' and ``reset'' scenarios), see Table 1 in Paper 1. The overall C/O ratio utilised is 0.34. In Paper 1 it was found that if the disk is exposed to
ionising cosmic rays, efficient production of \ce{CO2} ice and \ce{O2}
gas and ice takes place at intermediate disk radii, whereas \ce{CH4} gas
is converted to CO and \ce{CO2}.  Also, if full chemical reset into
atoms occurs en route into the disk midplane (e.g. via a shock or a
luminosity outburst), then the abundances of dominant volatiles can
differ from interstellar values by more than an order of magnitude
\citep[see also][]{drozdovskaya2016}. Most striking is the depletion
of \ce{H2O} ice advancing enhanced gas-phase CO and \ce{O2}. This
difference is larger for the case where cosmic rays are excluded from the ``reset'' scenario. \citet{helling2014} also found cosmic rays to be of importance to chemistry that alters the C/O ratio in the midplane. Their results, however, were different from those in Paper 1, due to differences in the grain-surface chemistry utilised.

Throughout this work, whenever distinguishing between gas-phase and icy species is neccessary, an ``i'' in front of a molecule involved in a reaction will denote that the molecule is icy, and a ``g'' in front will denote a gas-phase molecule. For reference, an iceline is the midplane radius inside of which a given chemical species is in the gas phase, and outside of which the species is in the ice.

\section{Results}
\label{results}

\begin{figure*}
\subfigure{\includegraphics[width=0.5\textwidth]{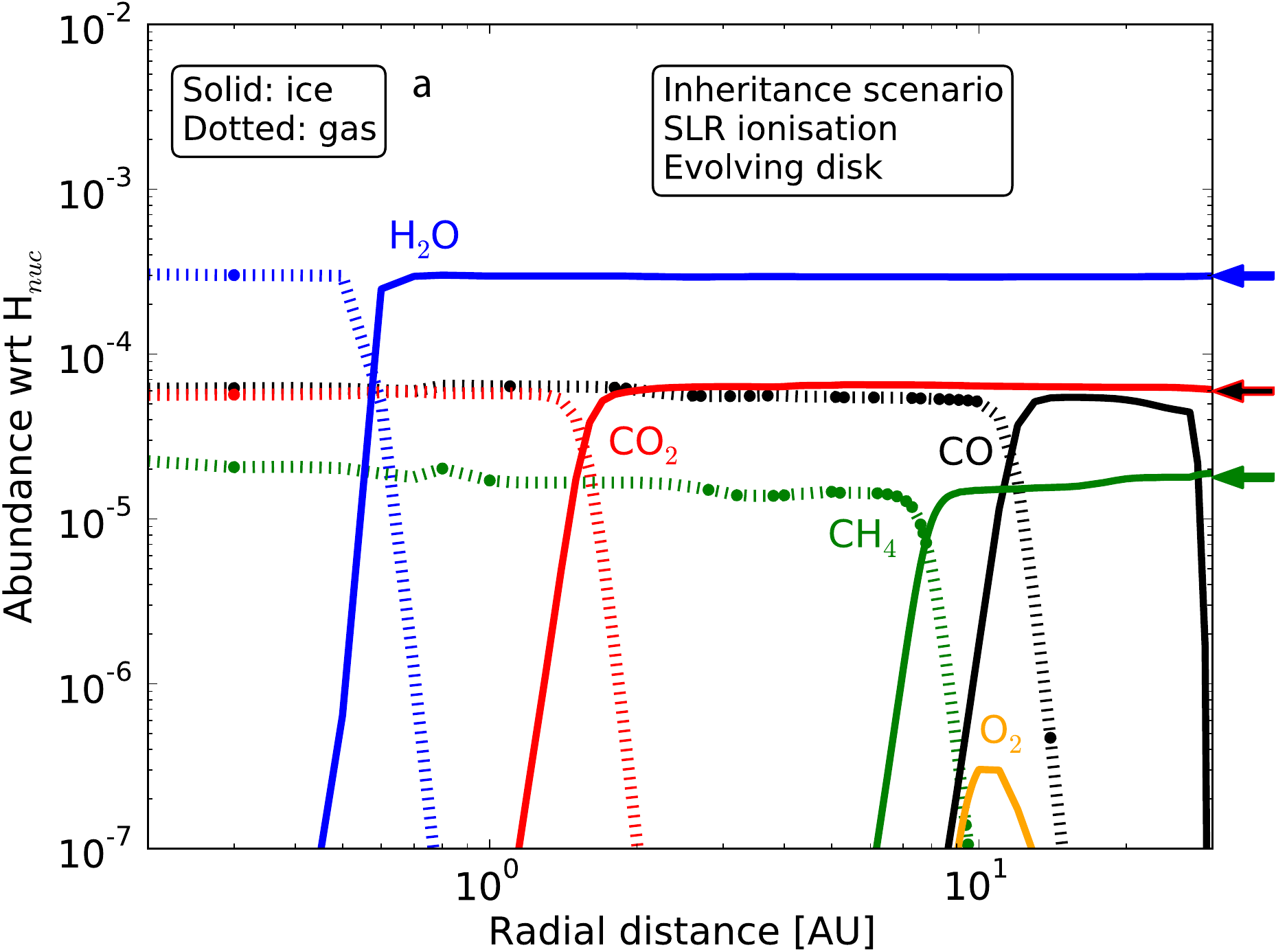}}
\subfigure{\includegraphics[width=0.5\textwidth]{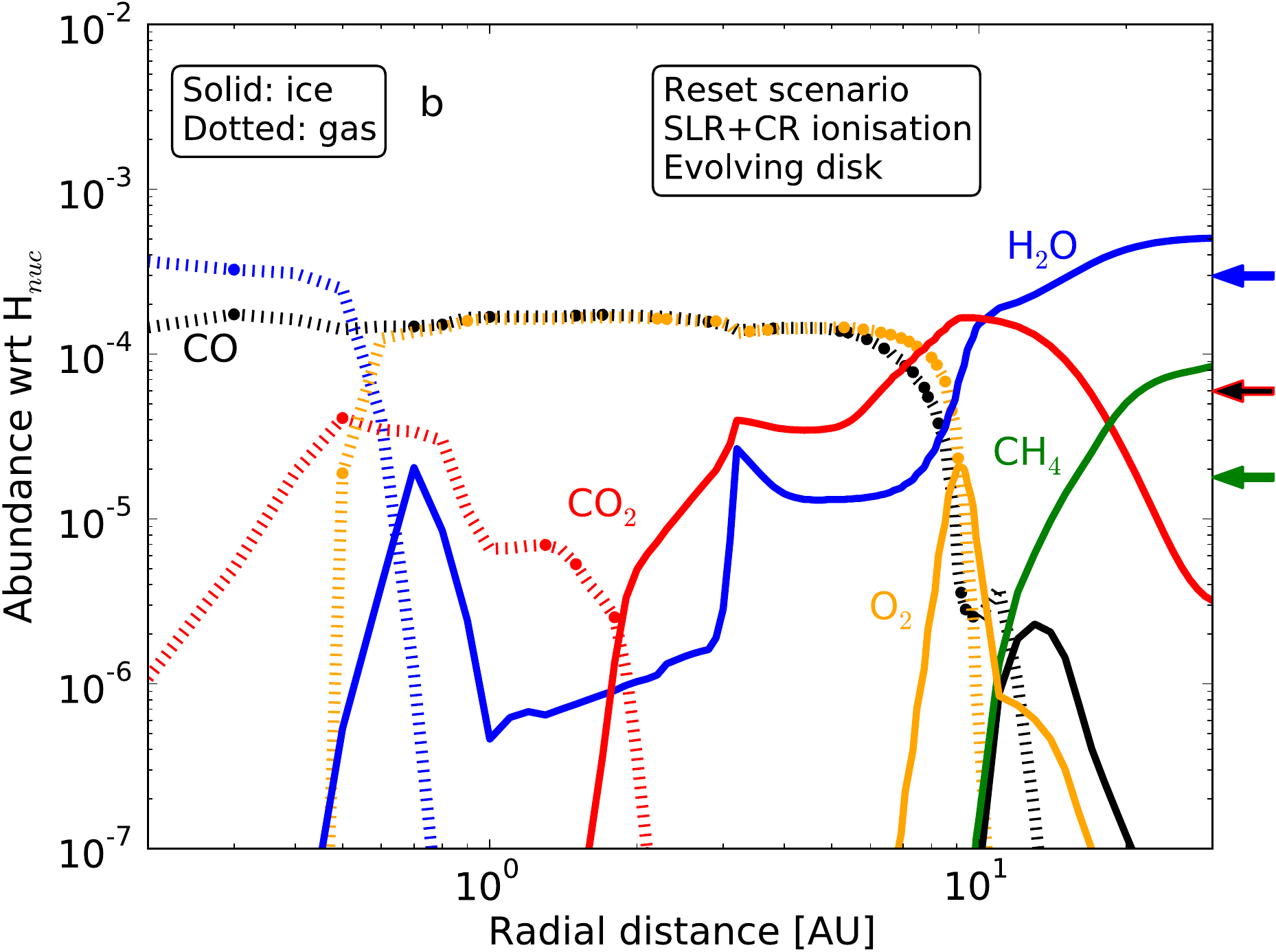}}\\
\subfigure{\includegraphics[width=0.5\textwidth]{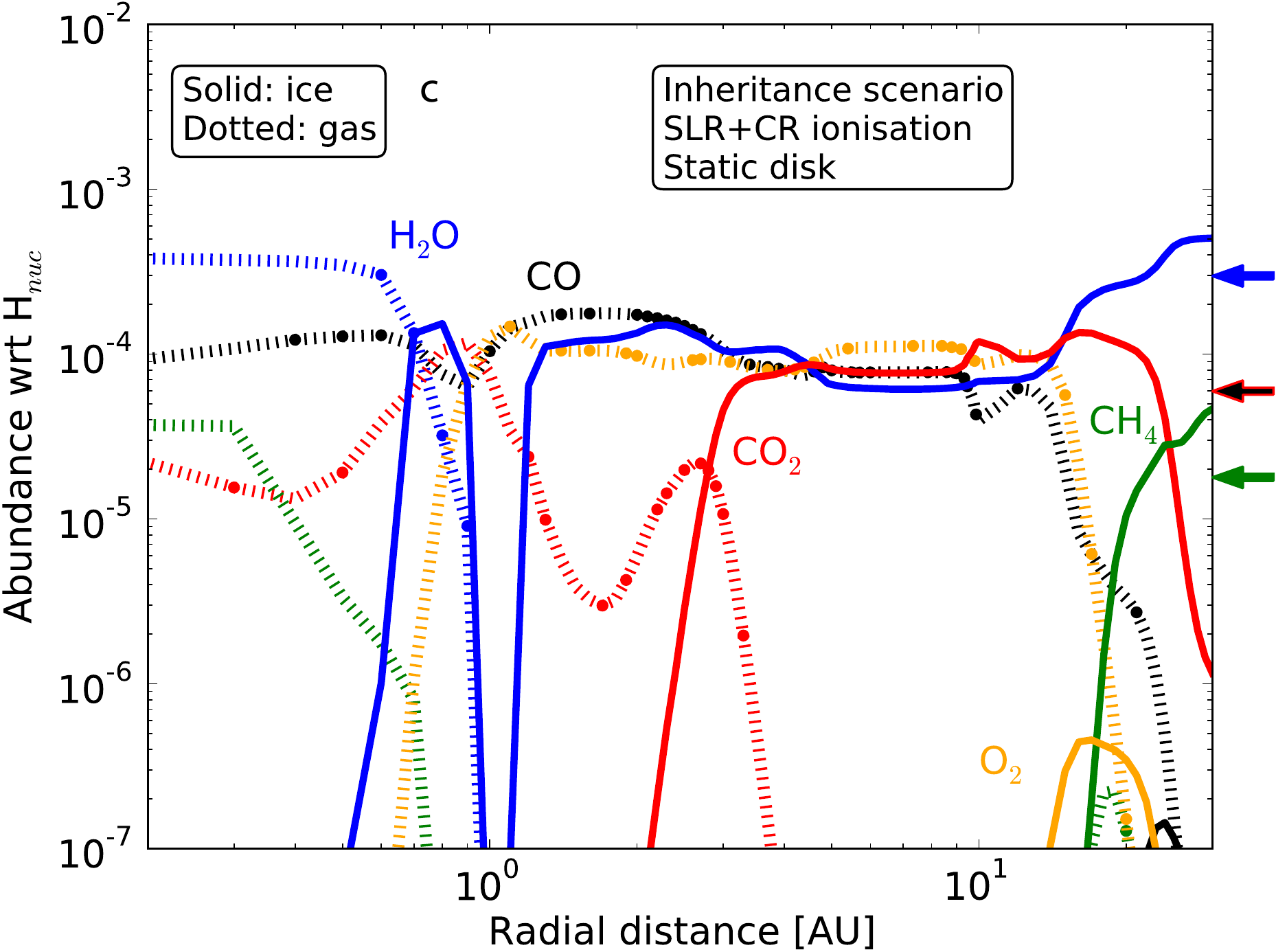}}
\subfigure{\includegraphics[width=0.5\textwidth]{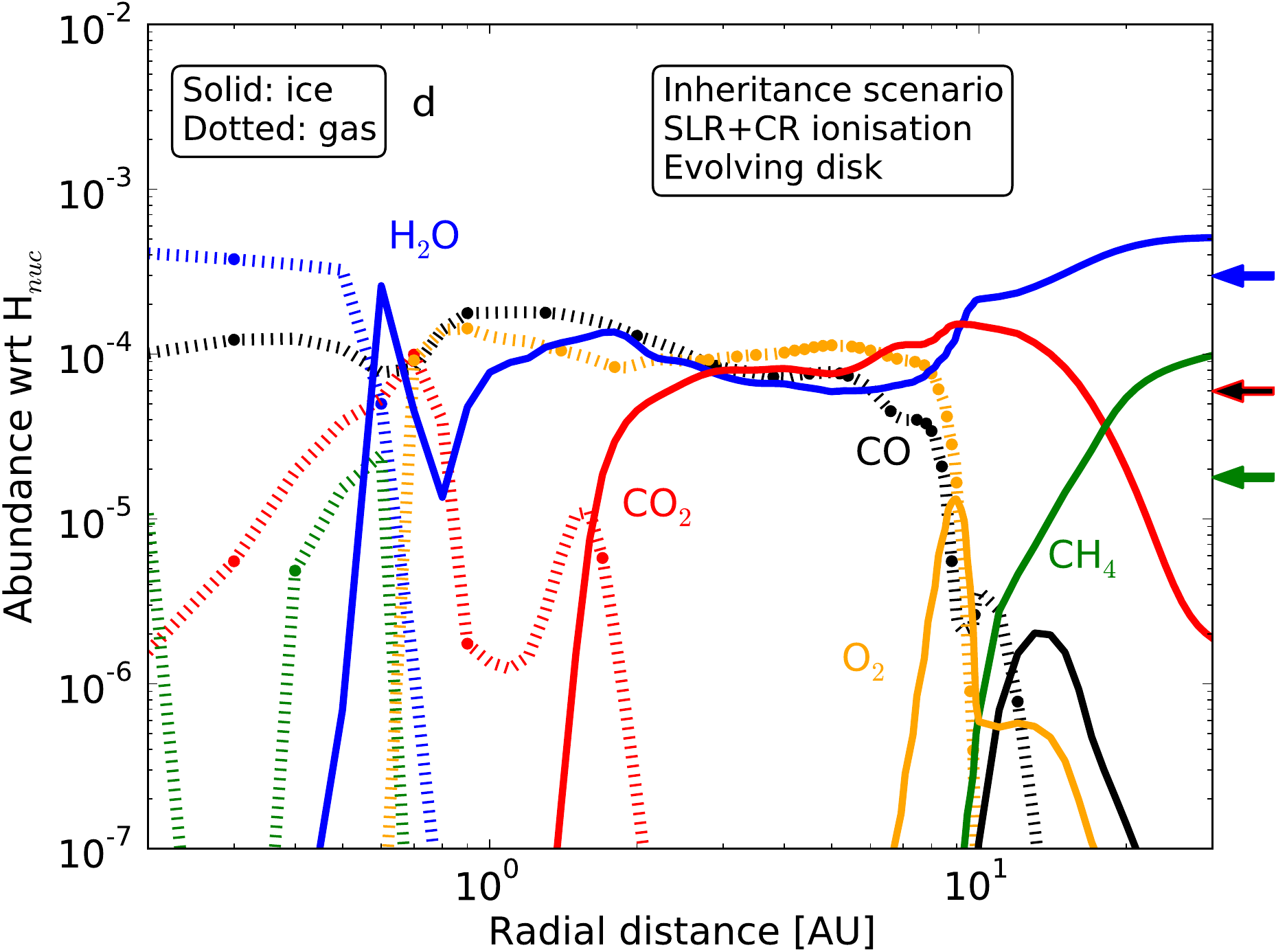}}\\
\caption{Abundances after 7 Myr evolution for four different model setups. Panel a: inheritance scenario with low ionisation and evolving disk structure. Panel b: reset scenario, high ionisation and evolving disk structure. Panel c: Inheritance scenario with high ionisation and static (unchanging) disk structure. Panel d: Inheritance scenario with high ionisation and evolving disk structure. The blue, red/black and green arrows to the right of each panel indicate the initial abundances assumed for \ce{H2O}, CO/\ce{CO2} and \ce{CH4}, respectively, in the inheritance scenario.}
\label{inh_vs_reset}
\end{figure*}

\begin{figure*}
\subfigure{\includegraphics[width=0.5\textwidth]{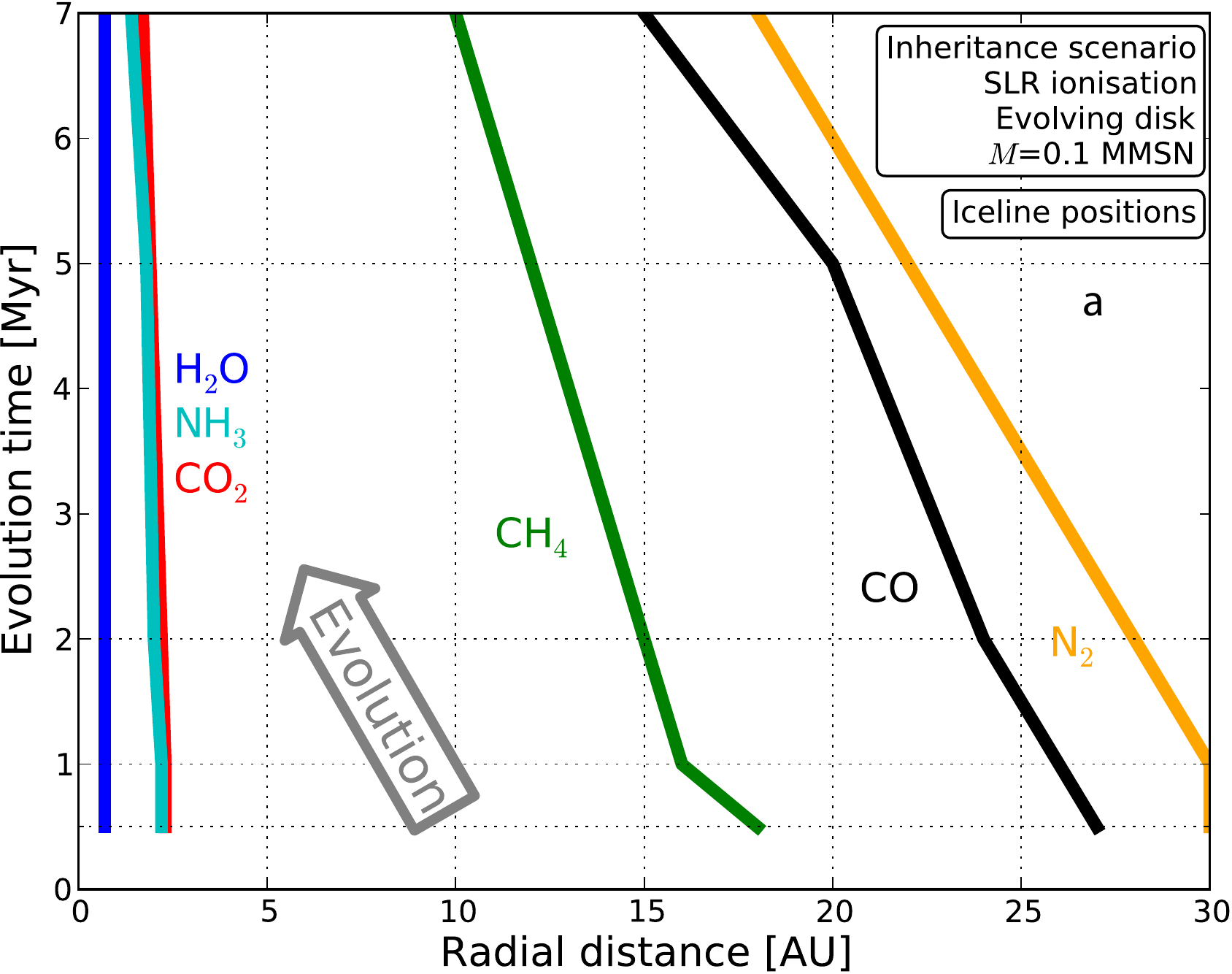}}
\subfigure{\includegraphics[width=0.5\textwidth]{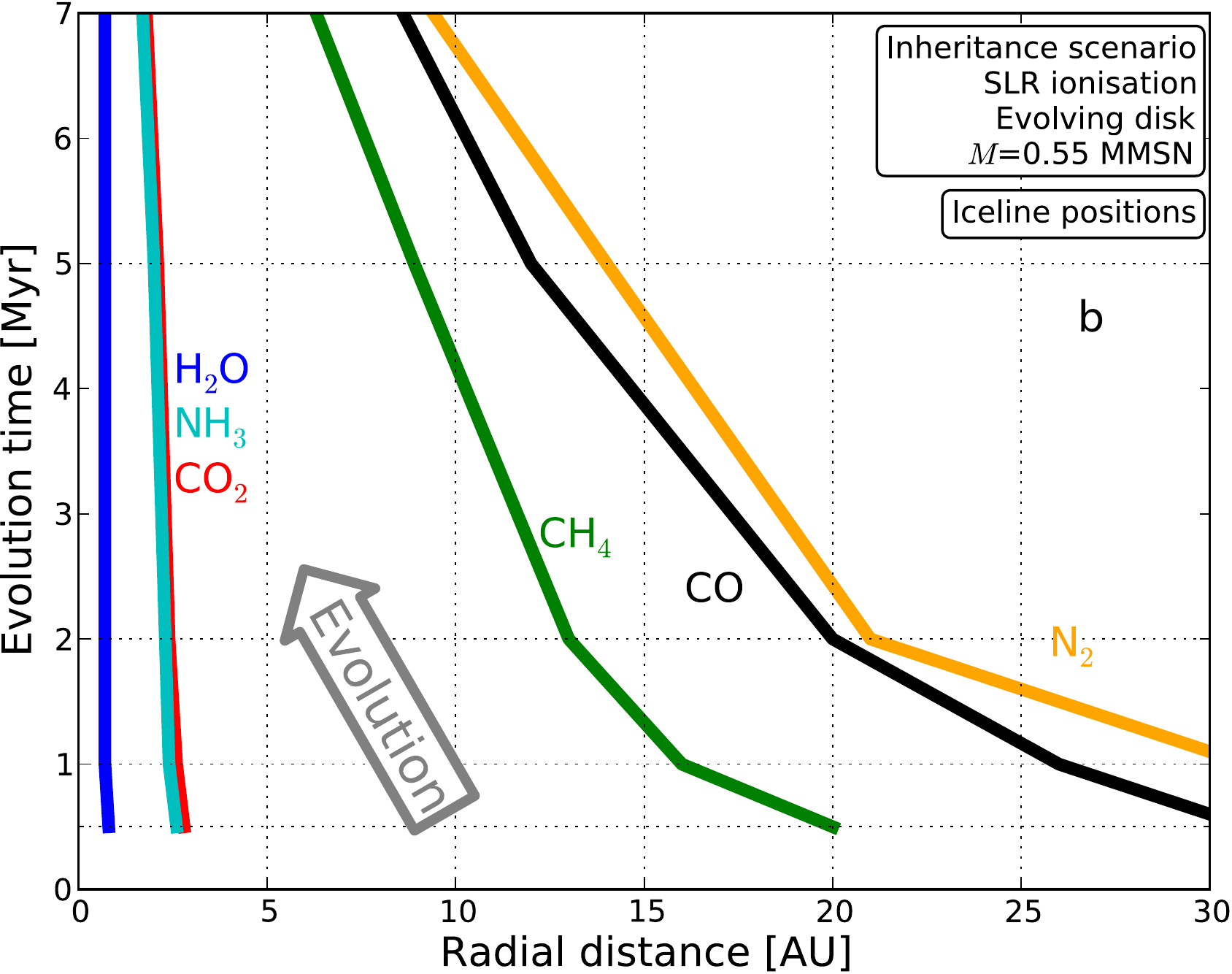}}\\
\caption{Evolving iceline positions for main volatiles, for 0.1 MMSN (left) and 0.55 MMSN (right) evolving disks}
\label{evol_icelines}
\end{figure*}

Fig. \ref{inh_vs_reset} shows abundances as a function of radius for selected volatile species in gas and ice after 7 Myr evolution, for four different model setups. As indicated in the boxes in each panel, the model setups explore a range of static and evolving disks, for the inheritance versus reset scenarios, and at both low and high ionisation levels.

\subsection{Static versus evolving disk: shifting icelines}
\label{disc_icelines}

In Paper 1 it was found that given low ionisation and inherited initial chemical abundances, chemical evolution was insignificant on the timescale of 1 Myr. Fig. \ref{inh_vs_reset}a shows the same case, but now for a physically evolving disk structure (the 0.1 MMSN disk), as described in Section \ref{methods}, and for up to 7 Myr. It is evident from Fig. \ref{inh_vs_reset}a that the chemistry in the midplane remains defined by the iceline positions, and that chemical processing has caused no significant chemical changes. This conclusion holds for the 0.55 MMSN evolving disk structure as well. Hence, the conclusion from Paper 1, that in the case of inherited abundances and low ionisation level, chemical evolution is insignificant, can be extended to a timescale of 7 Myr for these cases.

The main effect of utilising an evolving disk structure instead of a static one is that the icelines shift inwards with time. This in evident when comparing Fig. \ref{inh_vs_reset}c and \ref{inh_vs_reset}d, which both feature the inheritance scenario at high ionisation, but for a static and an evolving disk structure, respectively. For Fig. \ref{inh_vs_reset}d the icelines are shifted inwards in time, whereas for \ref{inh_vs_reset}c the icelines are static corresponding to the static temperature structure. The additional effects of adopting an evolving disk structure instead of a static structure are the retention of \ce{H2O} ice around 0.7 AU in panel d and 1-2 orders of magnitude lower \ce{CH4} gas abundance in the inner disk. The iceline positions for the disks are determined from Fig. \ref{inh_vs_reset} as the maximum distance $R$ from the star at which a volatile is more abundant in gas 
than in ice. Fig. \ref{evol_icelines} features the locations 
of the thermal icelines of selected volatile species as functions of time. The solid profiles indicate the movement 
of the icelines over time. Notice that the axes scales are linear in Fig. \ref{evol_icelines}.


For \ce{H2O}, \ce{NH3}, and \ce{CO2} in the 0.1 MMSN disk, the icelines shift by 30-35\% between 0.5 and 7 Myr of evolution 
(e.g., from 0.7 to 0.5 AU for \ce{H2O}). For \ce{CH4}, CO, and \ce{N2} the 
icelines are shifted inwards by $\sim$60\%. The larger shifts for the latter, 
more volatile species are due to the larger decreases in temperature in the outer 
disk, due to the decreasing slopes of the temperature profiles with time (see 
Fig.~\ref{low_mass_phys}a). For the 0.55 MMSN disk, the icelines shift farther than for the 0.1 MMSN disk. 
This is due to the steeper temperature profile for the 0.55 MMSN disk, as can be 
seen when comparing top left panels in Figs. \ref{low_mass_phys} and \ref{high_mass_phys}. The temperature drops by $\sim$50\% from 0 to 2 Myr at 10 AU in the 0.55 MMSN disk, versus a temperature drop by $\sim$20\% in the 0.1 MMSN disk over the same timeframe and radius.


\subsection{Timescales of chemical changes}
\label{disc_timescales}

\begin{figure*}
\subfigure{\includegraphics[width=0.5\textwidth]{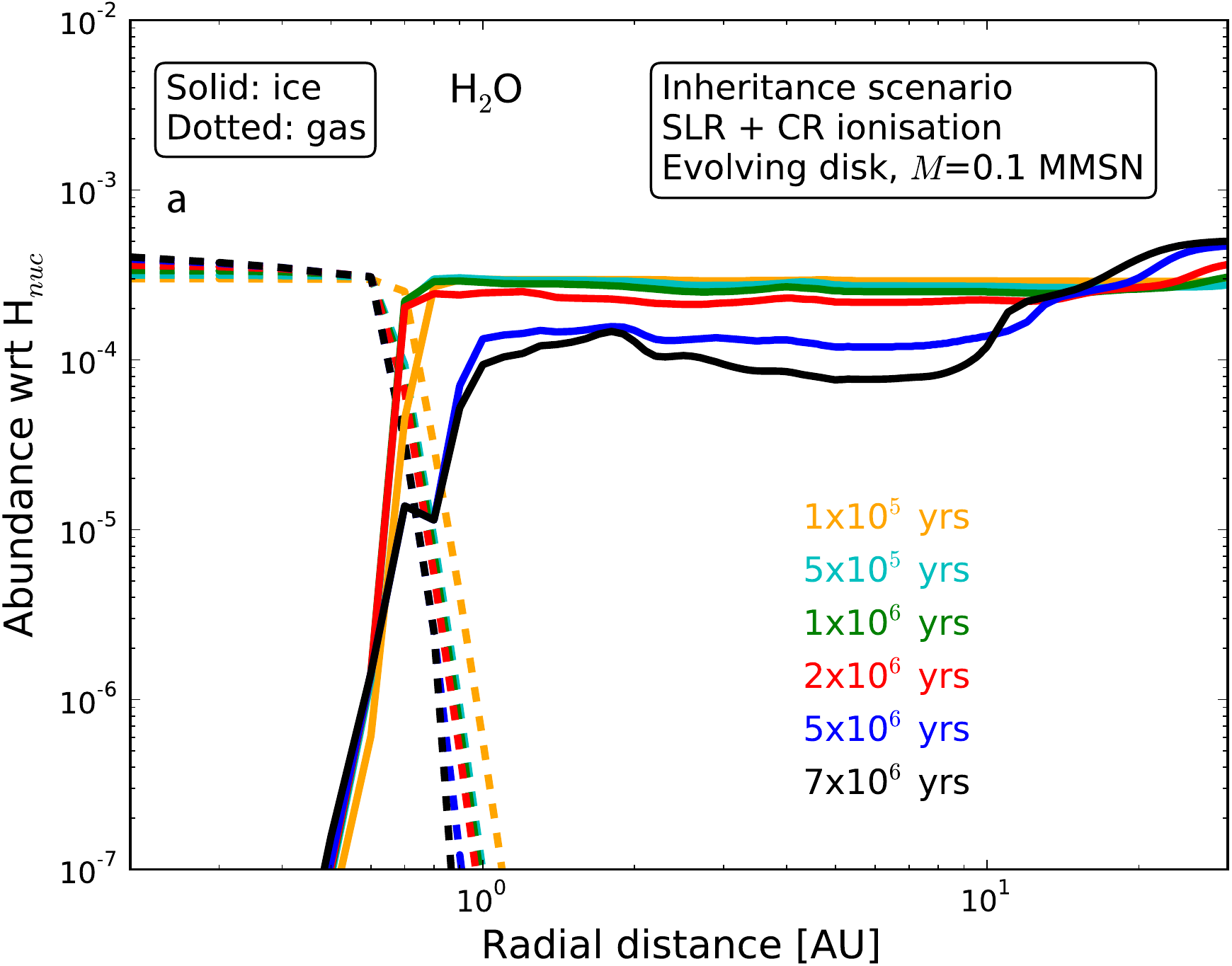}}
\subfigure{\includegraphics[width=0.5\textwidth]{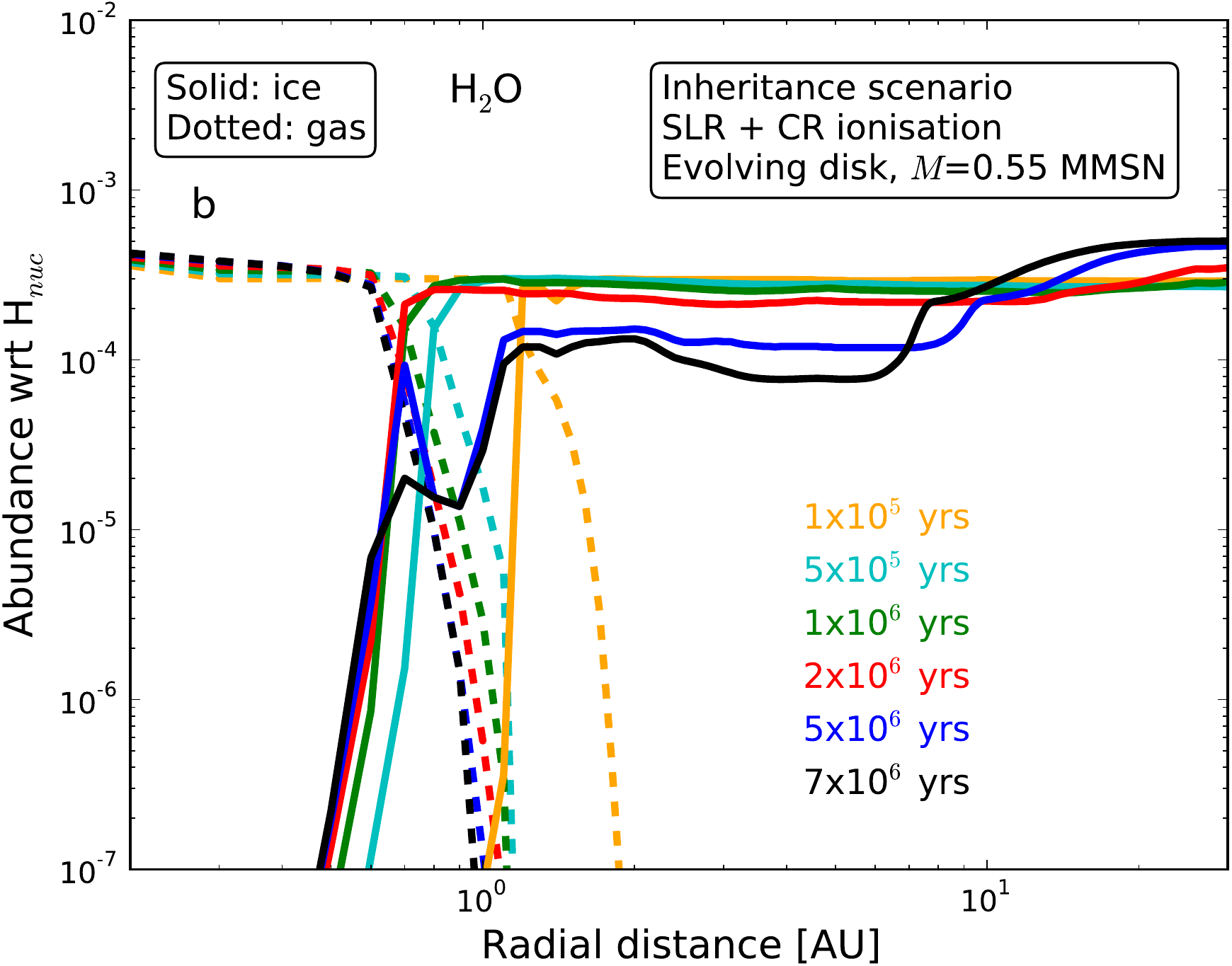}}\\
\subfigure{\includegraphics[width=0.5\textwidth]{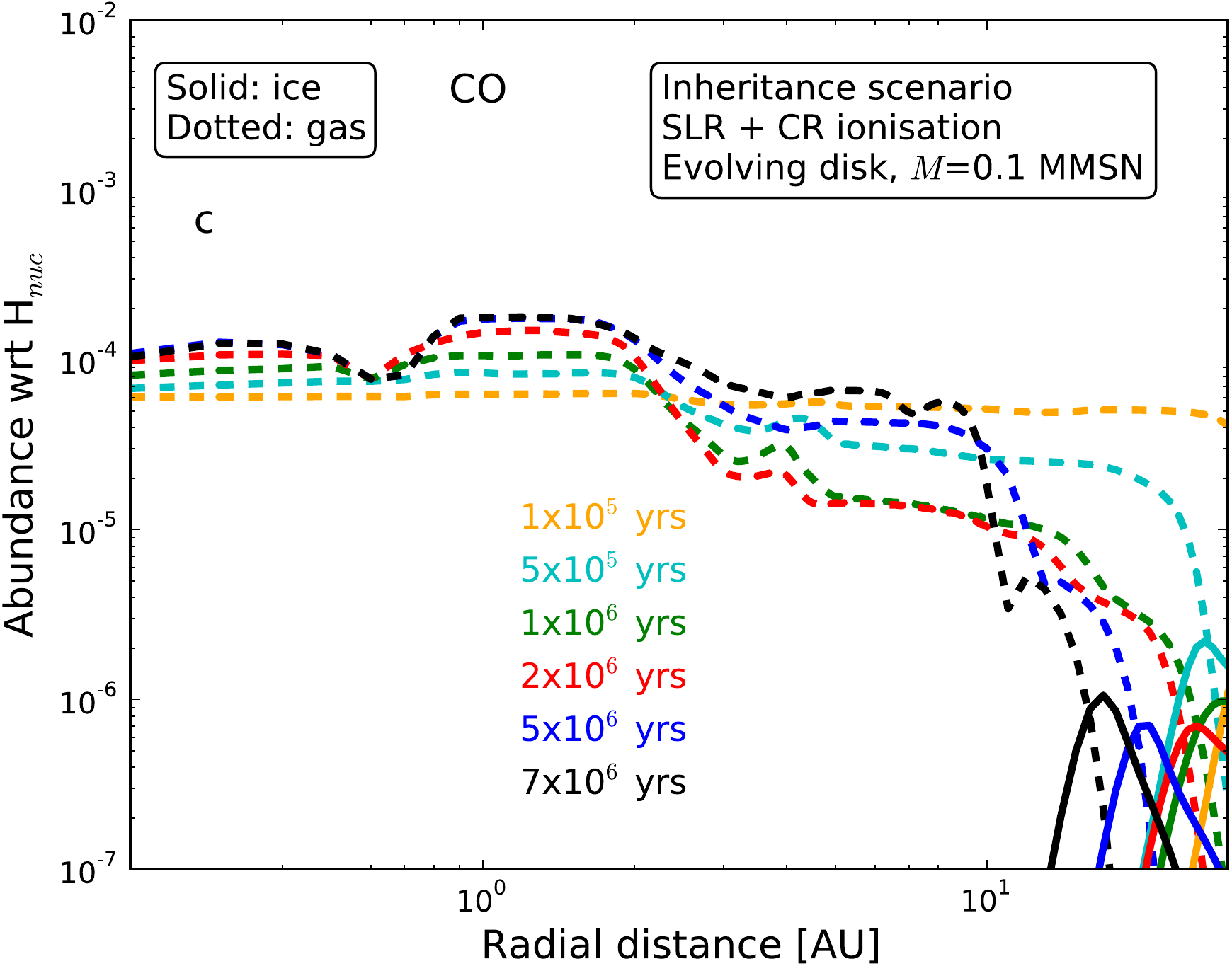}}
\subfigure{\includegraphics[width=0.5\textwidth]{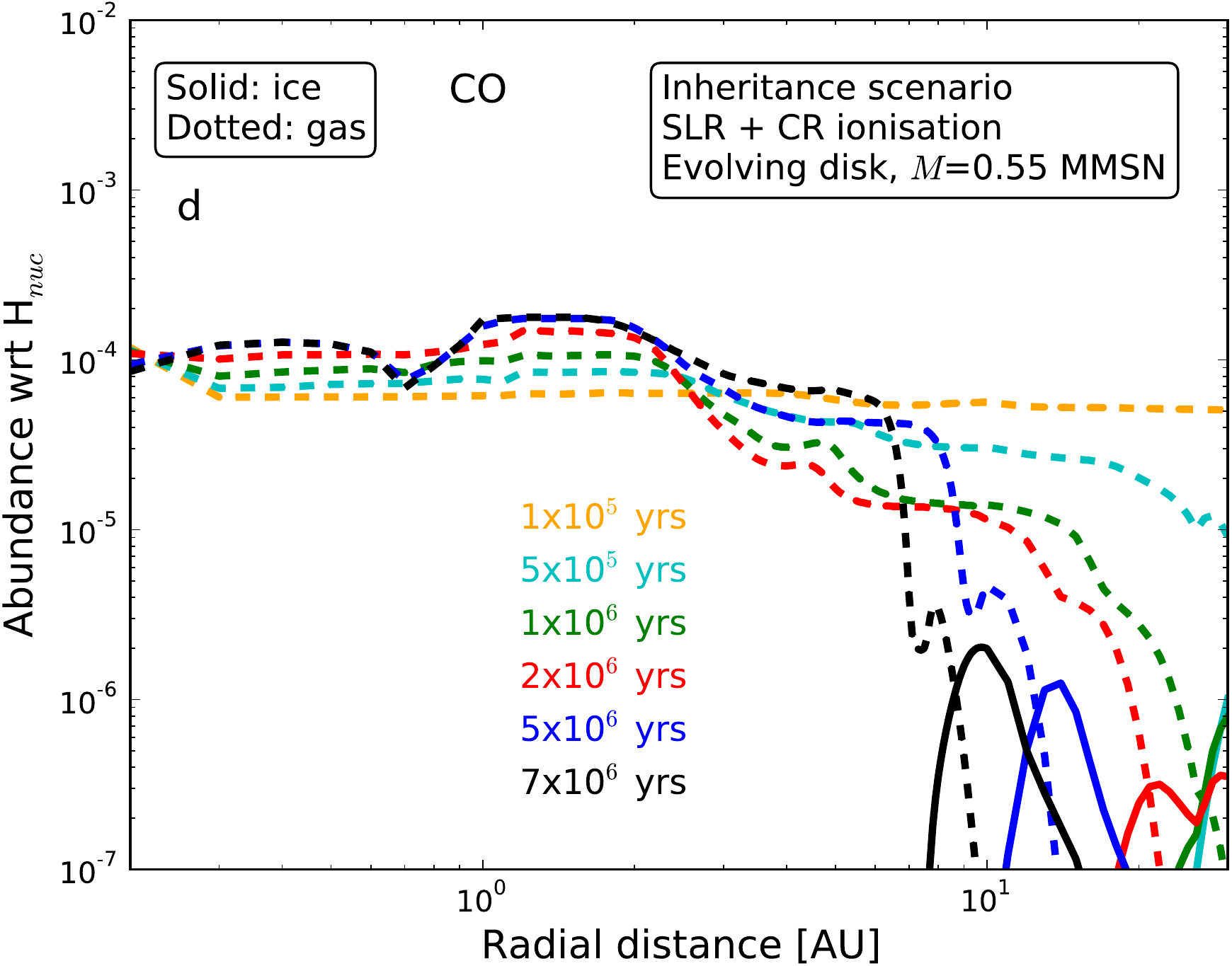}}\\
\subfigure{\includegraphics[width=0.5\textwidth]{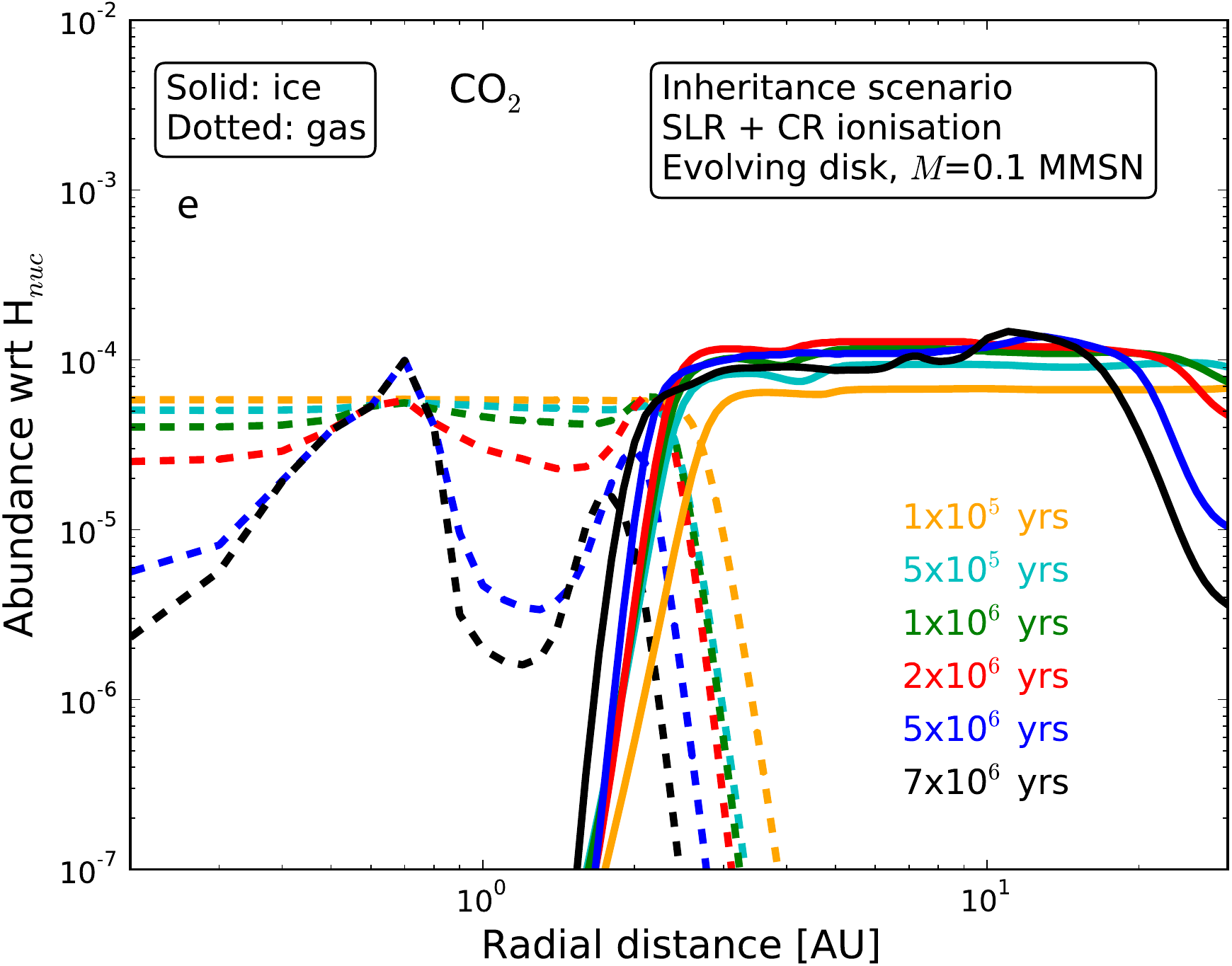}}
\subfigure{\includegraphics[width=0.5\textwidth]{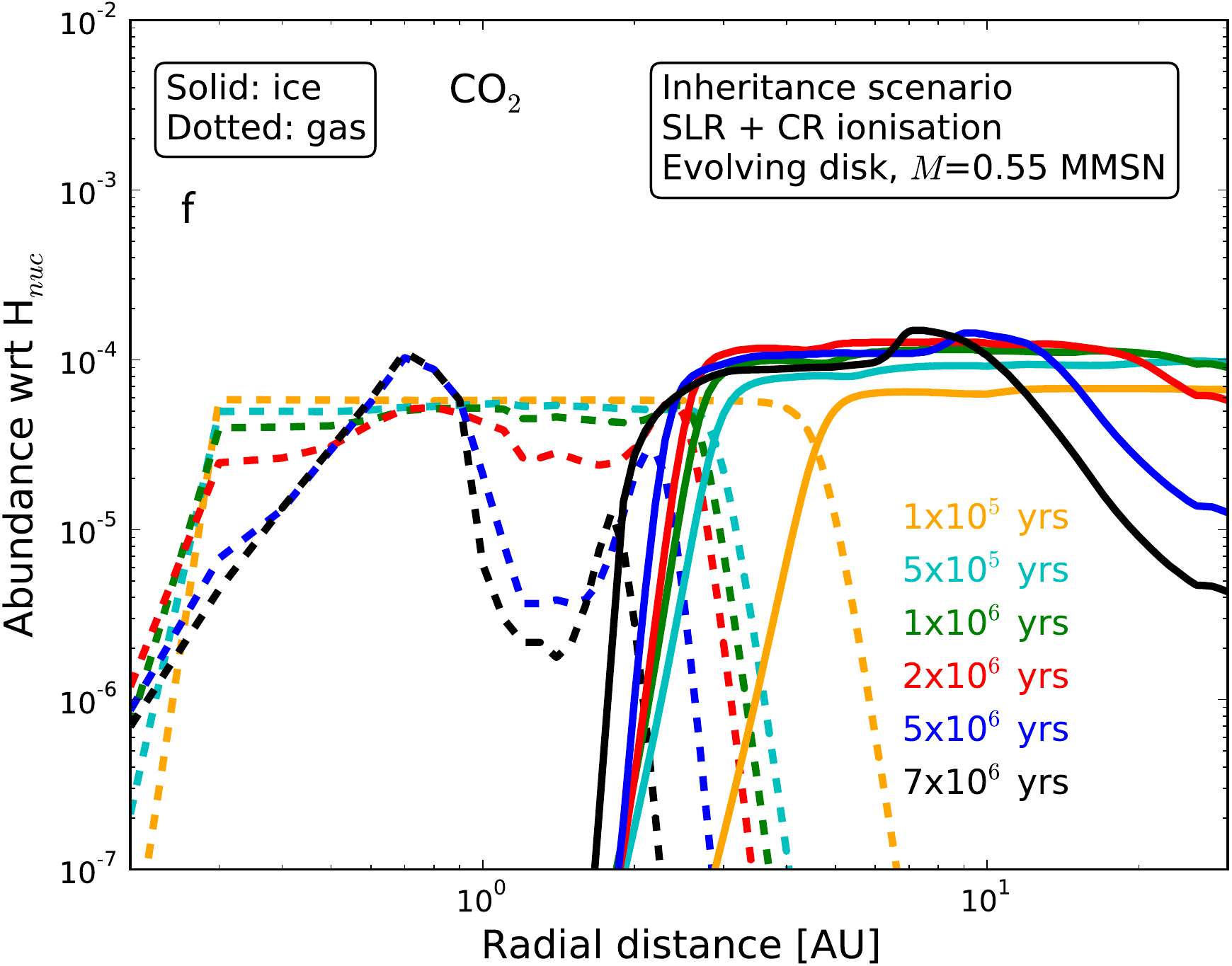}}\\
\caption{Abundances of \ce{H2O}, CO and \ce{CO2} gas (dashed) and ice (solid) 
as a function of midplane radius at five timesteps: 0.5, 1, 2, 5, and 7~Myr. 
The left-hand and right-hand panels show the 0.1 MMSN and 0.55 MMSN disk results, respectively.}
\label{low_vs_highmass1}
\end{figure*}

\begin{figure*}
\subfigure{\includegraphics[width=0.5\textwidth]{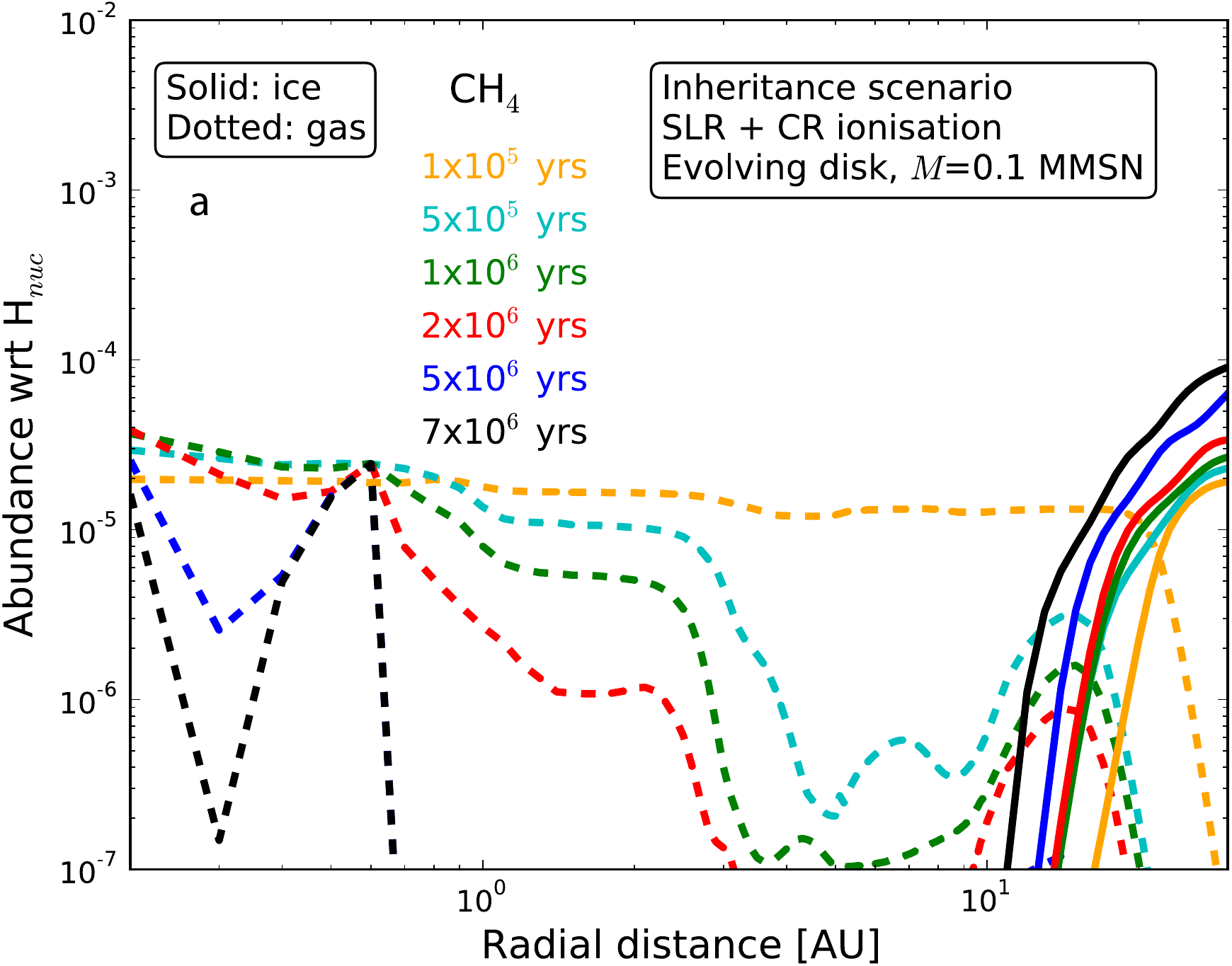}}
\subfigure{\includegraphics[width=0.5\textwidth]{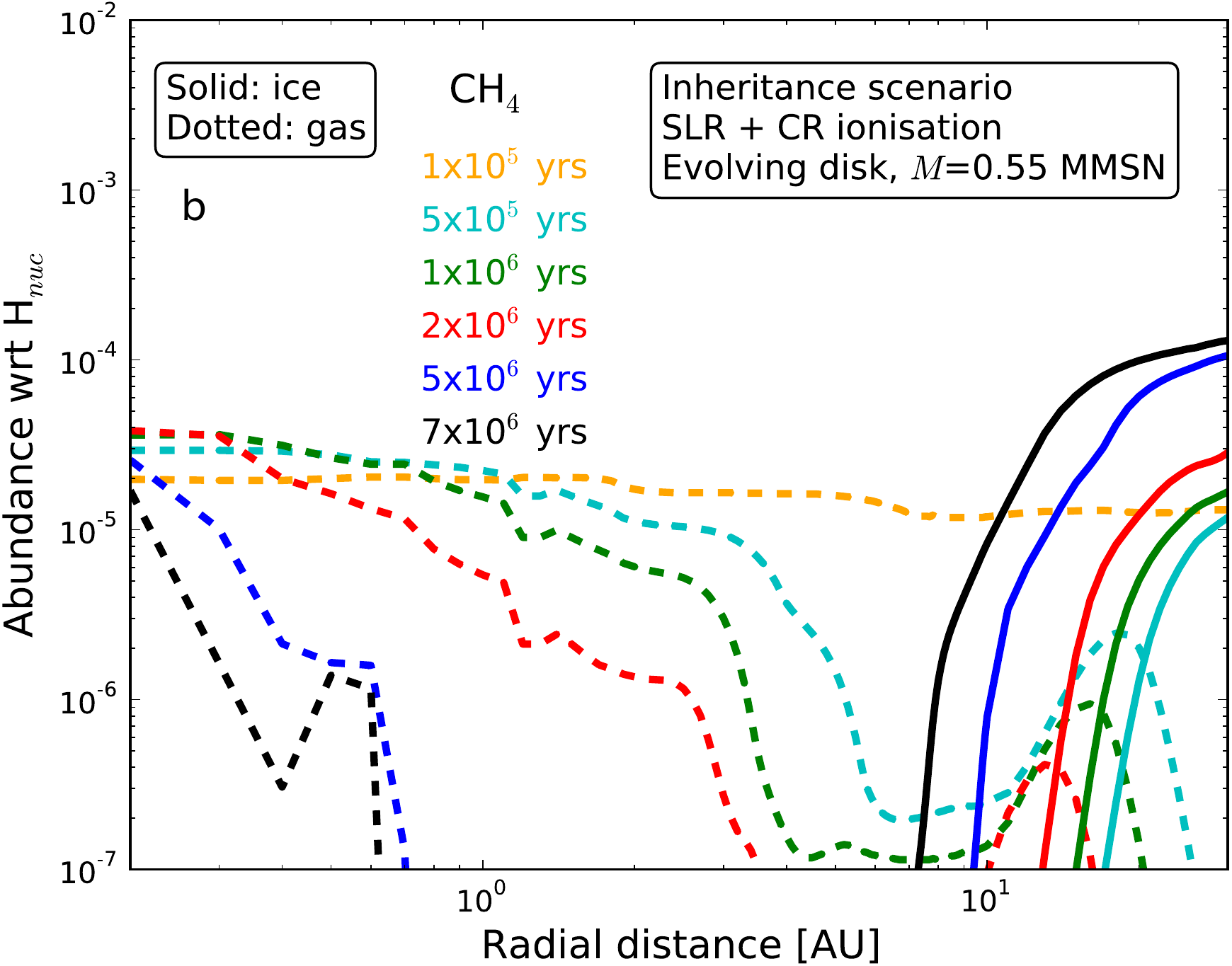}}\\
\subfigure{\includegraphics[width=0.5\textwidth]{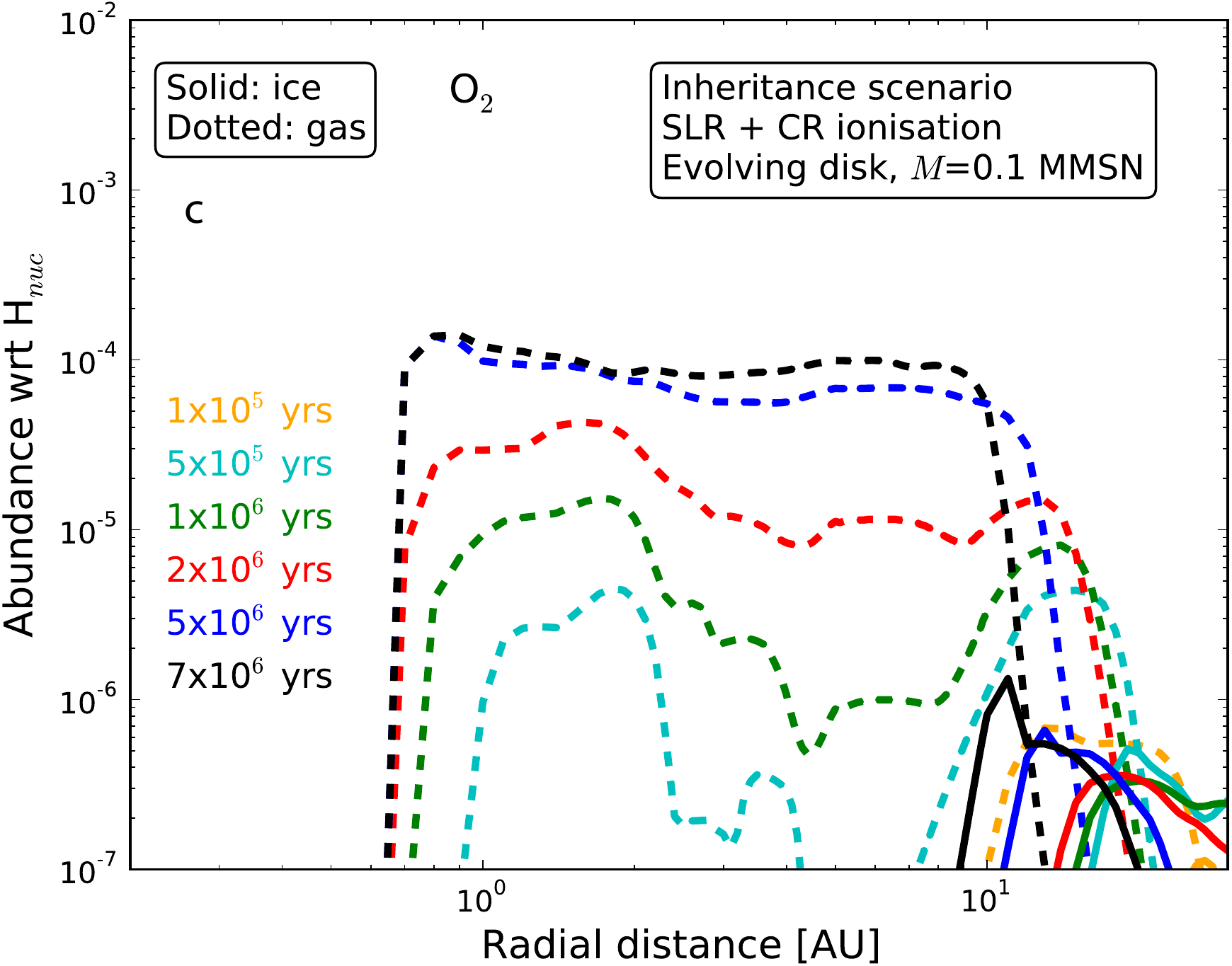}}
\subfigure{\includegraphics[width=0.5\textwidth]{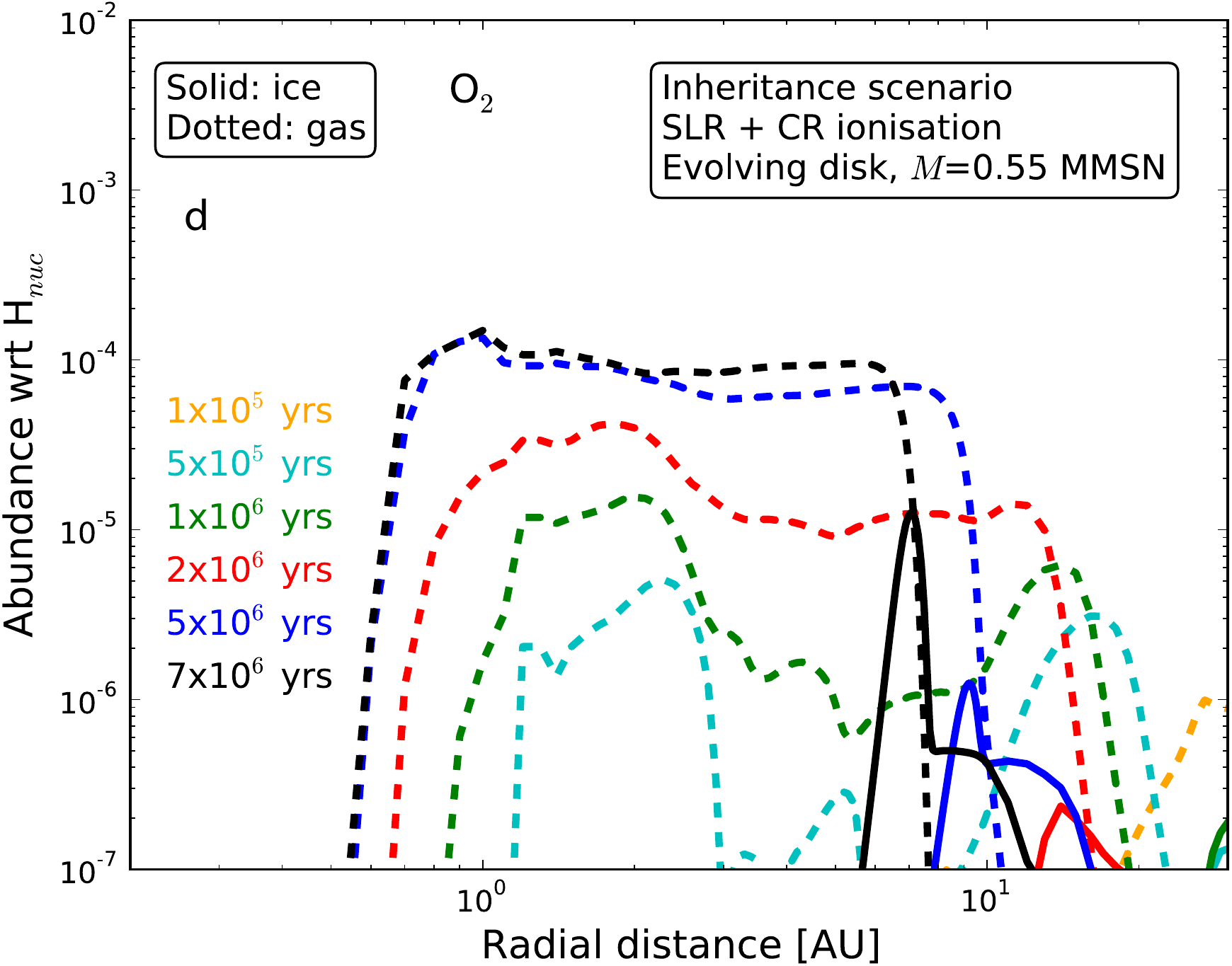}}\\

\caption{Same as Fig~\ref{low_vs_highmass1}, but for \ce{CH4} and \ce{O2}.}
\label{low_vs_highmass2}
\end{figure*}

Figs.~\ref{low_vs_highmass1} and \ref{low_vs_highmass2} show
abundances for five key volatiles as function of midplane 
radius at six timesteps: 0.1, 0.5, 1, 2, 5 and 7 Myr, for the ``inheritance'' scenario model at high ionisation. The left panels are for the 0.1 MMSN disk, and the right panels are for the 0.55 MMSN disk.
A time of 0.1~Myr was chosen as the first timestep because it was found in Paper 1 that for the inheritance scenario no chemical changes had taken place yet at this evolution time. An age of 0.5 Myr is 
representative of the timescale corresponding to the end of the 
embedded phase of star formation, when the protoplanetary disk is revealed. 
This was also found in Paper 1 to be the chemical evolution timescale 
for the high ionisation case. 
In that work, 1~Myr was considered the full evolution
time: here, we consider additional and later timesteps at 2, 5, and 7~Myr.  
The first is chosen as it is the median age of
T~Tauri stars and represents the lifetime of the gas-rich protoplanetary disk. 
A time of 5~Myr represents an aged disk like TW Hya.
By 7~Myr the gaseous disk is thought to have fully dissipated resulting 
in a gas-poor debris disk \citep[see e.g.][and Fig. \ref{evol_diskmass}]{fedele2010}.

As seen in Figs. \ref{low_vs_highmass1} and \ref{low_vs_highmass2}, chemistry is continuously ongoing throughout the evolutionary time of 7 Myr. The chemical abundance changes over 7 Myr vary from factors of a few (e.g. a 3 times decrease in \ce{H2O} ice abundance at 10 AU in Fig \ref{low_vs_highmass1}a to orders of magnitude (e.g. a \ce{CH4} gas depletion at 5 AU in \ref{low_vs_highmass2}a, resulting in a significantly different chemical ``picture'' from the beginning to the end of the evolution. For all species considered in these plots, the shifting icelines can be seen as an inward shift with time of the radius where the gas and ice profiles cross each other.
	

Overall, with the exceptions of \ce{H2O}, CO and \ce{CO2} inside 0.4 AU to be discussed in Section \ref{low_high}, the trends are similar for the 0.1 MMSN and 0.55 MMSN disks. Thus, in the following discussion only the chemical evolution of the 0.1 MMSN disk case is considered.

\begin{figure*}
\subfigure{\includegraphics[width=0.5\textwidth]{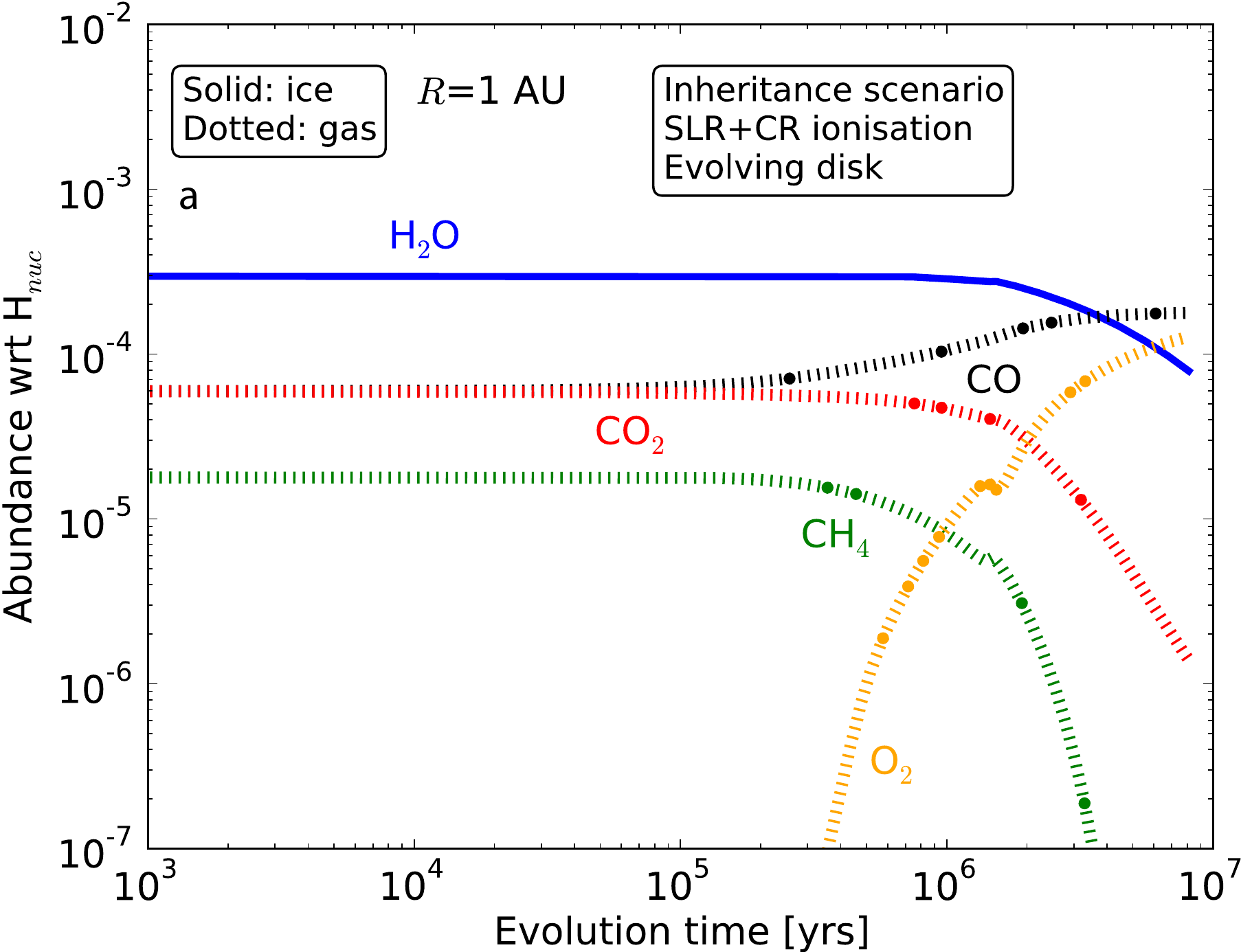}}
\subfigure{\includegraphics[width=0.5\textwidth]{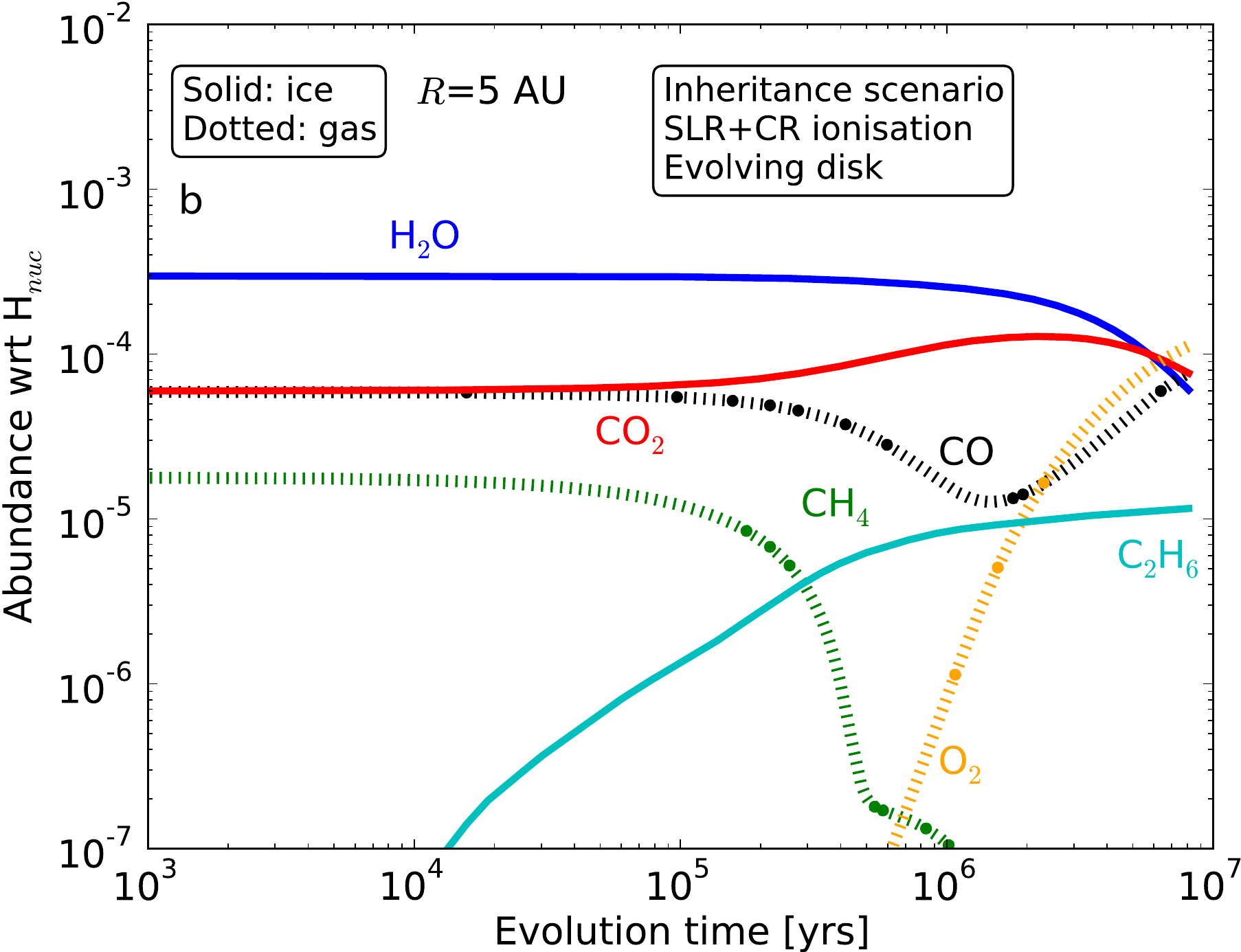}}\\
\subfigure{\includegraphics[width=0.5\textwidth]{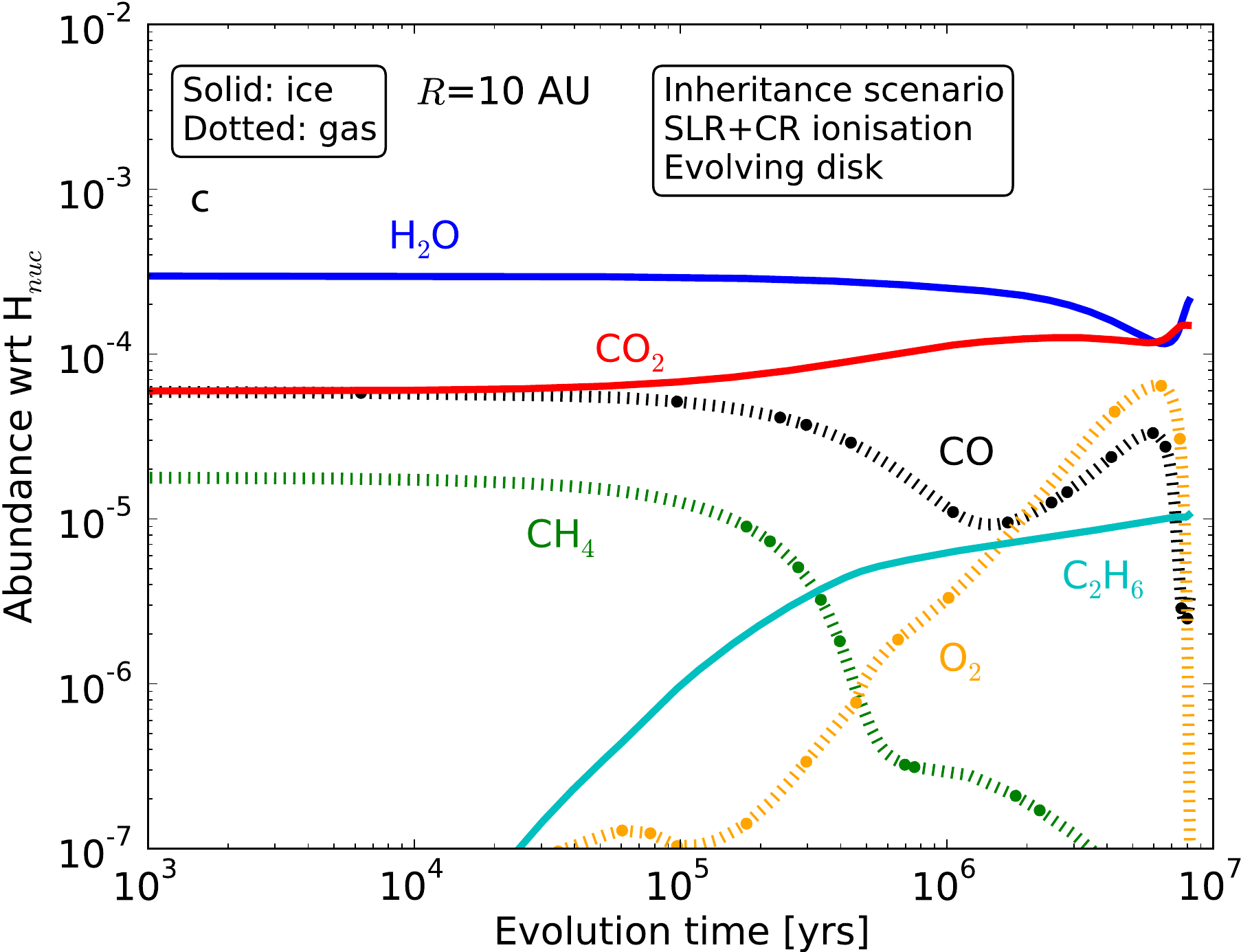}}
\subfigure{\includegraphics[width=0.5\textwidth]{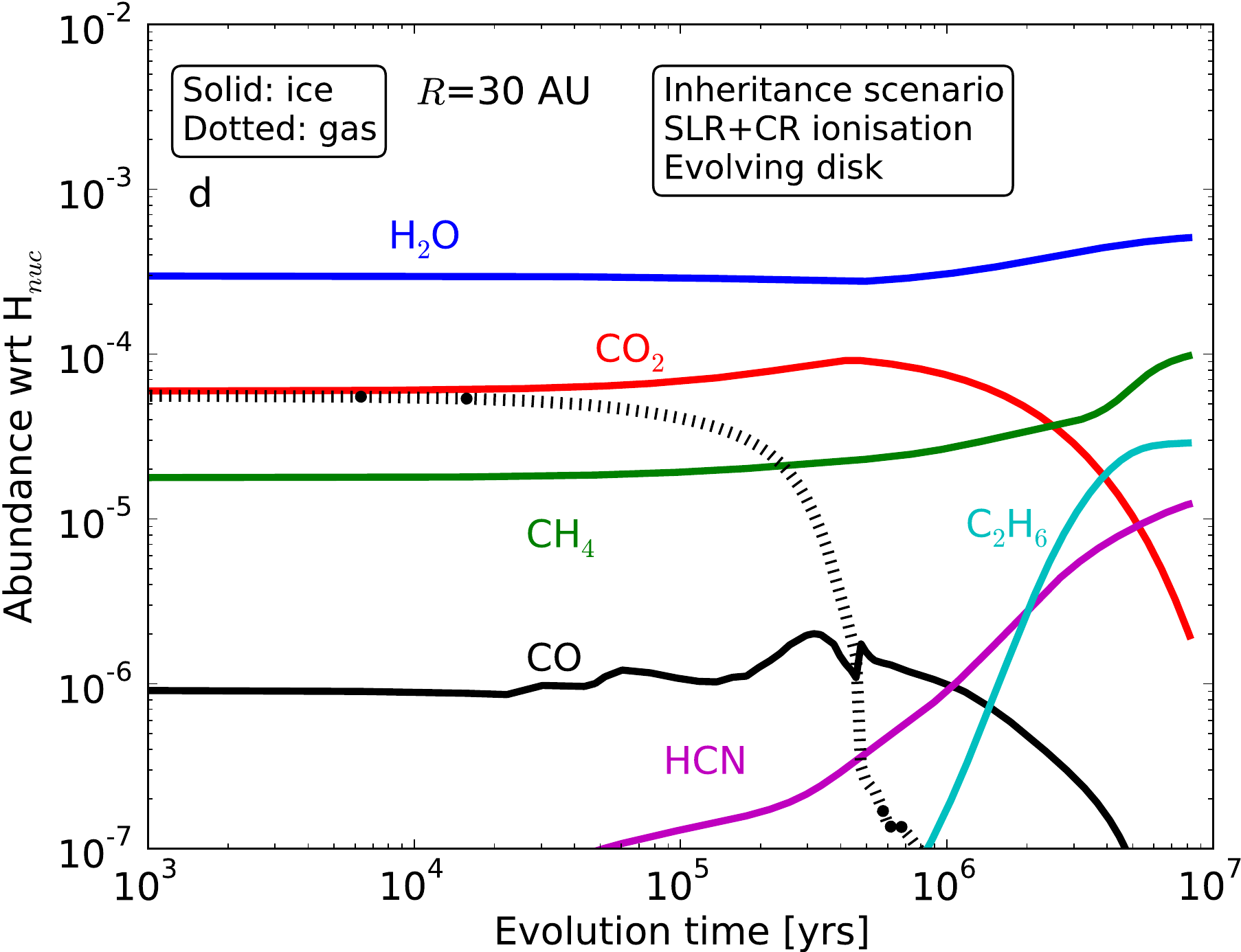}}\\
\caption{Time evolution of selected volatiles, in gas as ice at 1, 5, 10 and 30 AU, for the 0.1 MMSN disk with high ionisation and inheritance.}
\label{evolutions_lowmass}
\end{figure*}

To visualize the main chemical changes more clearly,
Fig.~\ref{evolutions_lowmass} highlights the evolution of key species
with time at each of four chosen radial positions: 1, 5, 10 and 30 AU. At 1 AU large changes happen between 1 and 7 Myr. The abundances
  of CO and \ce{O2} gas seem to converge at the end of the evolution,
  indicating that a steady state situation may be reached where these
  two species are the main carbon and oxygen reservoirs. \ce{H2O} ice is still dropping in abundance,
  thus, if evolution were to continue for a longer time, a
  further decrease in \ce{H2O} ice abundance would likely be seen.

At 5 AU, no steady state seems to have been reached after 7 Myr. CO and \ce{O2}
  gas and \ce{H2O} and \ce{CO2} ice have near equal abundances to within a
  factor of two, and a steady increase in \ce{C2H6} ice abundance is
  seen throughout the evolution reaching a final abundance of $\sim
  10^{-5}$, making it the third most abundant carrier of elemental carbon, accounting
  for $\sim 15$\% of the carbon. \ce{CH4} gas is depleted due to reactions with \ce{He+}, which is a consequence of the high ionisation level, as also mentioned in Paper 1. A "reverse" evolution of CO gas and
  \ce{CO2} ice is also evident, taking place at about the same
  time (between $\sim$1.6-5 Myr), indicating interdependency between
  production/destruction of CO and \ce{CO2}. The decreasing abundance of
  CO gas and increasing abundance of \ce{CO2} ice from $\sim$0.1 Myr to
  $\sim$2 Myr can be explained by CO gas colliding with the grains, and subsequently
  reacting fast with OH (faster than CO can desorp) to produce \ce{CO2} on grain surfaces, as
  also described in Paper 1. The ``reverse'' evolution for CO and \ce{CO2}, after 2 to 3 Myr, is due to the increasing midplane ionisation level over time (see Fig. \ref{low_mass_phys}d). For both a static and an evolving disk, the continuous CR ionisation of the midplane causes an increasing CR-induced photon fluence there (note that the fluence increases faster for the evolving disk due to the evolving ionisation level). This increasing fluence amplifies the photodissociation of \ce{CO2} ice into CO and O on the grains, which starts to dominate over the \ce{CO2} production reaction \ce{iCO + iOH -> iCO2 + iH}. The result is a decrease in \ce{CO2} ice and an increase in CO gas after 2-3 Myr. Thus, for an evolving disk, and the physical conditions considered here, the \ce{CO2} ice abundance reaches a maximum. The more diffuse the midplane becomes over time, the higher the ionisation contribution from CRs stays accordingly, and the shorter time it takes for the CO and \ce{CO2} evolutions to invert, in what will now be referred to as an ``abundance inversion''.
  
Going out to 10 AU, the evolution is very similar to that at 5 AU. The main
  difference is the rapid decrease in \ce{O2} and CO gas abundances at the
  end of the evolution, which is due to the shifting of the \ce{O2}
  iceline from outside to inside of 10 AU (see
  Fig. \ref{evol_icelines}), and the destruction of CO gas. The latter is caused by the rapid conversion of CO into \ce{CO2} ice mentioned above. These changes to \ce{O2} and CO gas at 10 AU
  remove them from the gas phase. An effect of this
  freeze-out is also the extra availability of \ce{O2} and CO on the grain
  surfaces which both react to form \ce{H2O} and \ce{CO2} ices, which are seen
  to increase in abundances simultaneously with the freeze-out of CO and \ce{O2}.

Out at 30 AU it is cold enough for all volatiles, except CO, to be frozen out initially. As seen in Fig. \ref{evolutions_lowmass},
  within the abundance limits, CO gas and ice are co-existing because
  $R$=30 AU is just inside the initial
  thermal CO iceline, with the ice abundance accounting for a few
  percent of the total CO abundance. The CO gas abundance decreases
  after $\sim$0.4 Myr, at which point the midplane has
  cooled enough for the CO iceline to move inwards of 30 AU, after which
  the CO ice becomes more abundant than the gas. However, the CO ice 
  abundance after about 1 Myr is only a few percent that of
  the initial CO gas abundance, indicating that CO is processed on
  grain surfaces upon freeze-out. \ce{CO2} ice is also destroyed after 1 Myr
  of evolution, but \ce{CH4} ice is steadily produced to become the dominant
  carbon-bearing species after 7 Myr. 
  
  This production of \ce{CH4} ice at 30 AU is linked to
  the destruction of \ce{CH3OH} ice through photodissociation by CR-induced photons which takes place simultaneously after 1 Myr. \ce{CH3OH} ice is
destroyed on the grains inside and outside of the location of CO
iceline. It is not reproduced from hydrogenation of CO because CO is
being converted into \ce{CO2}. Outside the CO iceline CO and \ce{CH3OH} are both being photodissociated on the grains caused by CR-induced photons, through the reactions,

\ce{iCH3OH ->[\gamma_{CR}] iCH3 + iOH}, and \ce{iCO ->[\gamma_{CR}] iO + iC}

This means low abundances of oxygen-containing complex organic molecules. In the cold outer midplane with temperatures down to around 15 K, the mobility of molecules on the grains is low, so hydrogenation reactions become dominant. \ce{iCH3 + iH -> iCH4} is causing the increase in \ce{CH4} ice abundance, with \ce{CH4} being the dominant carbon carrier after 7 Myr of evolution at 30 AU. Likewise, the hydrogenation of \ce{C2H4} ice and \ce{C2H5} ice is responsible for the production of \ce{C2H6} ice. \ce{H2O} ice is formed from hydrogenation of OH leftover from the photodissociation of \ce{CH3OH}, including the reaction: 

\ce{iCH3OH ->[\gamma_{CR}] iCH2 + iH2O}.

The abundance of \ce{C2H6} ice increases
with increasing radial distance. In addition to \ce{C2H6} ice, HCN ice is also abundantly produced at 30 AU, and these two species jointly account
  for 40\% of the elemental carbon, with \ce{CH4} accounting for an additional 50\%, and the remaining 10\% is in other molecules, such as \ce{CH3OH} (further discussed in Section \ref{gas_co}) by 7 Myr. \ce{H2O} ice accounts for more than 99\% of the elemental oxygen.



Overall, it is evident from these figures and discussion that chemical
evolution continues up to (at least) 7~Myr. That was also
a conclusion of \citet{kamp2013}. Under the conditions and assumptions
presented here, the majority of changes to the chemical abundances
take place between 0.5 and 7~Myr, and as in evident from Fig. \ref{low_vs_highmass1} and Fig. \ref{low_vs_highmass2}, the changes are significant (from factors of a few to orders of magnitude changes for \emph{all} species).

\subsection{Varying disk masses}
\label{low_high}

Figs. \ref{low_vs_highmass1} and \ref{low_vs_highmass2} present radial abundance distributions at different evolutionary timesteps for CO, \ce{CO2}, \ce{CH4}, \ce{H2O} and \ce{O2}. Left panels are for the low mass evolving disk structure. Right panels are for the high mass evolving disk structure.

Overall, when comparing the two disk masses for all five species it is seen that there are no significant differences in the abundance evolutions. However, the temperature structures are different, thus the icelines are located and shifted by different radial distances (see Section \ref{disc_icelines}). Only in the case for \ce{CO2} gas, is the evolution in the innermost disk (inside 0.4 AU) very different (by up to 3 orders of magnitude). This is due to the high temperatures and densities here, causing the abundance to be dictated by chemical equilibrium. This is more pronounced for the 0.55 MMSN disk, which is warmer in the inner tenth of an AU, than the 0.1 MMSN disk.

\subsection{Importance of initial abundances: inheritance vs reset}
\label{disc_initabun}

The choice of initial abundances was shown in Paper 1 to have a large 
effect on the final abundances after 1 Myr of evolution. In Fig. \ref{inh_vs_reset}, panels b and d show abundances after 7 Myr evolution 
for an evolving disk with a high level of ionisation, with panel b) assuming 
a reset scenario for the initial chemical abundance, and panel d) assuming 
the initial abundances to be inherited. Since both cases have assumed the 
same evolving physical structure, the icelines are identically located in 
both panels.

Inside the \ce{H2O} iceline at $\sim$0.5 AU, where all species are in the gas 
phase, similar abundance levels are seen for \ce{H2O}, CO and \ce{CO2}, in both scenarios. \ce{CH4} gas, on the contrary, is absent from 
the inner disk in the reset scenario. In relation to this absence, a 
destruction of \ce{CH4} gas is evident in the inheritance scenario as seen 
at 1 AU in Fig. \ref{evolutions_lowmass}a, which indicates that \ce{CH4} is being 
continuously destroyed in the gas phase, a destruction happening 
more slowly inside the \ce{H2O} iceline in Fig. \ref{inh_vs_reset}, than it does 
outside that iceline.

Redirecting focus from the inner to the outer disk, the abundances of the 
ices outside the \ce{O2} iceline at around 10 AU are very similar 
(at 30 AU \ce{CO2} is a factor of 2 different, \ce{CH4} is $\sim$10\% different, and \ce{H2O} is only 1\% different) when comparing the reset and the inheritance 
scenarios in Fig.~\ref{inh_vs_reset} b and d for chemical evolution by 7 Myr. Indeed, outside 10 AU the 
choice of initial abundances seems irrelevant for the resulting 
abundances, given a long enough evolution time. In other words: 
in this region, which is very relevant for comet formation, chemical steady state is 
close, and the fingerprint of the interstellar chemical abundances is 
erased if the gas disk lasts for 7 Myr.

The inner (mostly) gaseous midplane, and the outer (mostly) icy midplane 
produce similar abundance ratios independent of the choice of initial abundances. 
The intermediate region in Fig.~\ref{inh_vs_reset} b and d, roughly between the \ce{H2O} and \ce{O2} icelines at 0.5 AU 
and 10 AU, is quite different. The dominant gas phase species are CO and \ce{O2}, 
both at approximately the same abundance level within a factor of 2, in both 
the reset and the inheritance scenarios. \ce{H2O} and \ce{CO2} ices, however, 
are 1-2 orders of magnitude lower for the reset scenario than for the inheritance scenario. 
Looking into the time evolution of CO, \ce{O2}, \ce{H2O} and \ce{CO2} in the inheritance 
scenario with high ionisation at 5 AU shown in Fig. \ref{evolutions_lowmass}b, it is evident that none of these species have yet reached steady state (although it is close) by 7 Myr. 
Additional evolution time would therefore likely bring the abundances of 
\ce{H2O} and \ce{CO2} ices in the inheritance scenario further down, and CO and \ce{O2} gases further up, similar to the plot for the reset scenario in Fig. \ref{inh_vs_reset}b. 





\subsection{\ce{O2} as a significant disk midplane molecule}

Molecular oxygen is often not considered to be a dominant molecule in planet formation models. The detection of a high abundance of \ce{O2} ice to \ce{H2O} ice of 3.8+/-0.85\% on Comet 67P reported by \citet{bieler15} has triggered models to explain the origin of this molecular oxygen \citep[see][]{mousis2016,taquet2016}.  

In disk chemistry models by \citet{walsh15} \ce{O2} is abundantly produced in the intermediate layers of a disk. Here, the \ce{O2} evolution in the midplane is traced up to 7 Myr. Fig. \ref{low_vs_highmass2} shows abundance profiles for \ce{O2} at different timesteps for the ``inheritance'' scenario. It is evident that between the \ce{H2O} and \ce{O2} icelines, the abundance of \ce{O2} gas continuously increases with time. \ce{O2} is produced on the grain surfaces through the following reaction pathway:

\ce{iH2S ->[\rm{H}] iHS ->[\rm{O}] iSO ->[\rm{O}] iSO2 ->[\rm{H}] iO2},

\ce{O2} ice subsequently desorbs (if inside the \ce{O2} iceline), and steadily builds up a high abundance of $\sim$10$^{-4}$ wrt \ce{H2O} over 7 Myr evolution. The initial abundance of \ce{H2S} is 6$\times 10^{-6}$ (20 times lower than the \ce{O2} gas abundance by 7 Myr), and the HS molecule left over from the last step in the reaction chain above is recycled for conversion of more oxygen atoms into \ce{O2} via the second step above. Henceforth, the initial abundance of sulphur is seemingly irrelevant for the final abundance of \ce{O2}, since the sulphur merely acts as a catalyst for the conversion of oxygen atoms into molecular oxygen.


\subsection{Caveats of model assumptions}
\label{disc_caveats}

\subsubsection{Choice of grain-surface reaction parameters}

An important part of the chemical evolution presented here is caused by grain-surface chemistry. The importance of grain-surface chemistry was also highlighted in Paper 1, as well as in \citet{walsh14a}. However, the grain-surface reactions are sensitive to the choice of grain surface parameters. In this work, the barrier height for quantum tunnelling $b_{\rm{qt}}$ between grain surface sites has been set to $b_{\rm{qt}}=1$ \AA ng, and the ratio of diffusion energy $E_{\rm{diff}}$ between grain surface sites and molecular binding energy $E_{\rm{des}}$ has been set to 
$E_{\rm{diff}}/E_{\rm{des}}=0.5$. This is a conservative take where most grain-surface chemistry happens at moderate temperatures ($\gtrsim25$K). If $b_{\rm{qt}}$ was to be increased, the hydrogenation reactions would be slowed down, and vice versa for decreasing $b_{\rm{qt}}$. If $E_{\rm{diff}}/E_{\rm{des}}$ was to be decreased, then grain-surface radical-radical recombination could happen at lower temperatures. Such changes would affect chemical evolution, but an exploration of this parameter space is left for future work.

\subsubsection{Dependence on ionisation}

As highlighted in several previous studies, including Paper 1, \citet{aikawa96}, \citet{helling2014}, \citet{cleeves2014crex} and \citet{walsh15}, the chemical evolution taking place in protoplanetary disk midplanes is highly dependent on the degree of ionisation. The timescale of chemical evolution has been found in this work to scale inversely with the ionisation level. It is found that an ionisation level 10 times higher than that presented here will shorten the chemical evolution time scale 10 times, and vice versa.

In the situation where the midplane is significantly shielded from CR ionisation, e.g. by a high column density attenuating the CR particles, or by magnetic field effects \citep[see][]{cleeves13crex}, then no significant chemical evolution takes place over the disk lifetime of 7 Myr. In the case of complete exclusion of CRs, this is seen in Fig. \ref{inh_vs_reset}a, where the ionisation level is solely based on SLR decay products in the midplane. This ionisation level starts out between 10$^{-19}$ and 10$^{-18}$~s$^{-1}$, and drops orders of magnitude over the evolution time, as seen in Fig. \ref{low_mass_phys}c. Thus, for midplane ionisation levels of a few times 10$^{-19}$ s$^{-1}$ or lower, 7 Myr of evolution is insufficient time for the ionisation to cause significant chemical evolution. Along the same lines, if chemical steady state is to be achieved within 1 Myr of evolution, in the outer disk, then the ionisation level has be $\sim10^{-16}$s$^{-1}$ (about ten times higher than assumed for high ionisation in the outer disk here) or higher. The effects of X-ray ionisation on disk chemistry in planet-forming midplanes will be explored in future work.


\subsubsection{Choice of grain sizes and distribution}
\label{caveat_grains}

Like in Paper 1, the physical grain parameters have been kept constant throughout the evolution time. However, in real disks several processes  take place for grains during the lifetime of a disk, all of which could change the grain parameters adopted here. These processes include grain settling to the midplane, agglomeration and fragmentation, as well as grain radial drift. The physics of grain evolution has been studied extensively \citep[see e.g.][and references therein]{birnstiel2016,kataoka2014,okuzumi2012}.

One way of theoretically investigating the effects of changing dust parameters, is to simply adopt a different dust size as input for the models. Increasing the dust size will reduce the total surface area-to-mass ratio, prolonging the freeze-out and desorption timescales, and vice versa for decreasing dust size. However, the grain-surface chemical evolution of the ices will likely simply be slowed down for a reduced surface area-to-mass ratio, and sped up for an increased surface area-to-mass ratio. Thus, the same chemical effects might be seen, just on different timescales (see Section \ref{chem_model}). Recent work by \citet{facchini2017} found that the dust growth timescale in the disk midplane is significantly longer (3 orders of magnitude) than the timescale of freeze out and desorption, e.g. for CO. 

The dust growth timescale is, however, still shorter than the timescale for chemical evolution. Although this means that the grain size utilised for grain-surface reactions here in fact is not static (as assumed) over the chemical evolution time, the grain growth might not affect the chemistry that differently from what has been presented in this work: grain growth likely does not accumulate small compact spherical grains into larger and compact grains with a predictable change in grain surface areas. Rather the growth may lead to fluffy aggregates or grains with uneven surfaces \citep{blum2008}. Hence, the change to the grain surface area-to-mass ratio may not be significant. If grains are subsequently compactified resulting in a reduced grain surface area-to-mass ratio, it may simply lead to the aforementioned longer timescale for grain-surface reactions.


\section{Discussion}
\label{discussion}

\subsection{Iceline shifts: comparison to other studies}

The positions and movements of volatile icelines have been subject to much investigation recently, see \citet{piso2015}, \citet{alidib2017}, \citet{oberg2016}, and references therein. \citet{piso2015} modelled shifting iceline positions 
over 3 Myr assuming both radial drift of particles, and a viscous disk 
with gas accretion. In these models, they found shifted icelines of volatiles compared to a static 
disk situation. These shifts, however, are not caused by evolving temperature structures as is the case in the present work, but by radial drift of different sized particles. Thus, radial drift can also shift the icelines independently by setting different radii for desorption of icy grain mantles depending on the sizes of the grains. Despite these differences, both evolving temperatures and radial drift of particles act to shift the icelines inwards. \citet{piso2015} found that the \ce{H2O} iceline was shifted inwards by up to 40\%, whereas the icelines of \ce{CO2} and CO were shifted 
inwards by up to 60\% and 50\%, respectively. These shifts are comparable in size to those found in this work, but not caused by the same mechanism. Thus, the combination of different physical disk evolution effects could possibly result in even larger iceline shifts.

\subsection{What happens to gaseous CO?}
\label{gas_co}

The dominant ices in the outer disk midplane at the end of evolution are all stable molecules, for both choices of initial abundances. For the ``reset'' scenario this results from hydrogenation of carbon and oxygen atoms on the grain surfaces, whereas for the ``inheritance'' scenario, the oxygen and carbon atoms are mainly supplied from the CR-induced photodissociation of CO and \ce{CO2}.

CO gas has been studied extensively through observations of protoplanetary disks. As noted in the introduction several observational studies have found surprisingly weak CO emission from disks, more than can be accounted for by simple freeze-out and photodissociation. In particular for TW Hya, \citet{schwarz2016} reported depletion by two orders of magnitude through CO isotopologue observations, and \citet{zhang2017} recently determined the CO midplane iceline with an associated freeze-out temperature of $27^{+3}_{-2}$ K, higher than expected from the binding energy of pure CO ice ($\sim 21$K). \citet{zhang2017} attributed this higher freeze-out temperature to CO being frozen out onto a mix of \ce{H2O} and \ce{CO2} ices. In relation to these reported depletions, the findings for CO in this work are interesting. The outer icy disk here is not the only region lacking CO. A decrease in CO gas abundance is also seen to take place for the ``inheritance'' scenario for the region between the \ce{CO2} and the CO icelines in Fig. \ref{low_vs_highmass1}c around $\sim$3 AU, until 2 Myr of evolution. The effect of CO freezing out and subsequently reacting into \ce{CO2} and chemically converting into other species such as HCN and \ce{C2H6} means that CO gas is effectively reduced in abundance by up to 2 orders of magnitude, within the CO iceline. This reduction in CO gas, however, is only happening so long as the CR-induced photon fluence is low enough to not dissociate \ce{CO2} into CO on the grains, as discussed in Section \ref{disc_timescales}. The models show that chemical CO gas depletion in a disk midplane could therefore be a temporary effect. As discussed earlier, the timescale of CO depletion, which lasts until the ``abundance inversion'' takes place, will depend on the ionisation conditions, and thus the density, in the midplane, with a lower ionisation level prolonging the CO depletion timescale.

An additional large drop in CO gas abundance is seen starting by 5 Myr at $\sim$10 AU for the 0.1 MMSN disk, see Fig. \ref{low_vs_highmass1}c (the ``inheritance'' scenario).
Thus, if one was to determine the CO iceline from depletion of gas-phase CO isotopologues in an old disk like TW Hya, the desorption temperature for CO in this model could appear to be at $\sim$29 K \citep[similar to the freeze-out temperature for CO found by][]{zhang2017} at $\sim$10 AU, rather than at $\sim$21 K at 16 AU, for the 0.1 MMSN disk. This appearance of the CO iceline to be at a higher temperature and smaller radius than prescribed by the CO binding energy is due to the chemical depletion of CO gas, happening outside the \ce{O2} iceline, resulting in a ``fake'' CO iceline. A ``fake'' CO iceline was also seen in Paper 1, although that was different from the one here: in Paper 1 the ``fake'' iceline was seen outside the \ce{CO2} iceline by 1 Myr, thus at a smaller radius and at an earlier evolutionary stage. The ``fake'' iceline seen here coincides with the thermal iceline of \ce{O2}, and the reduction in CO gas abundance outside 10 AU is due to the freeze-out of \ce{O2}. \ce{O2} ice is hydrogenated into \ce{HO2} ice on the grains, and subsequently dissociated into OH ice. This OH, in turn, reacts with colliding CO to produce \ce{CO2} ice, thereby reducing the CO gas abundance. This effect will be further investigated in future work.

\begin{figure*}
\subfigure{\includegraphics[width=0.345\textwidth]{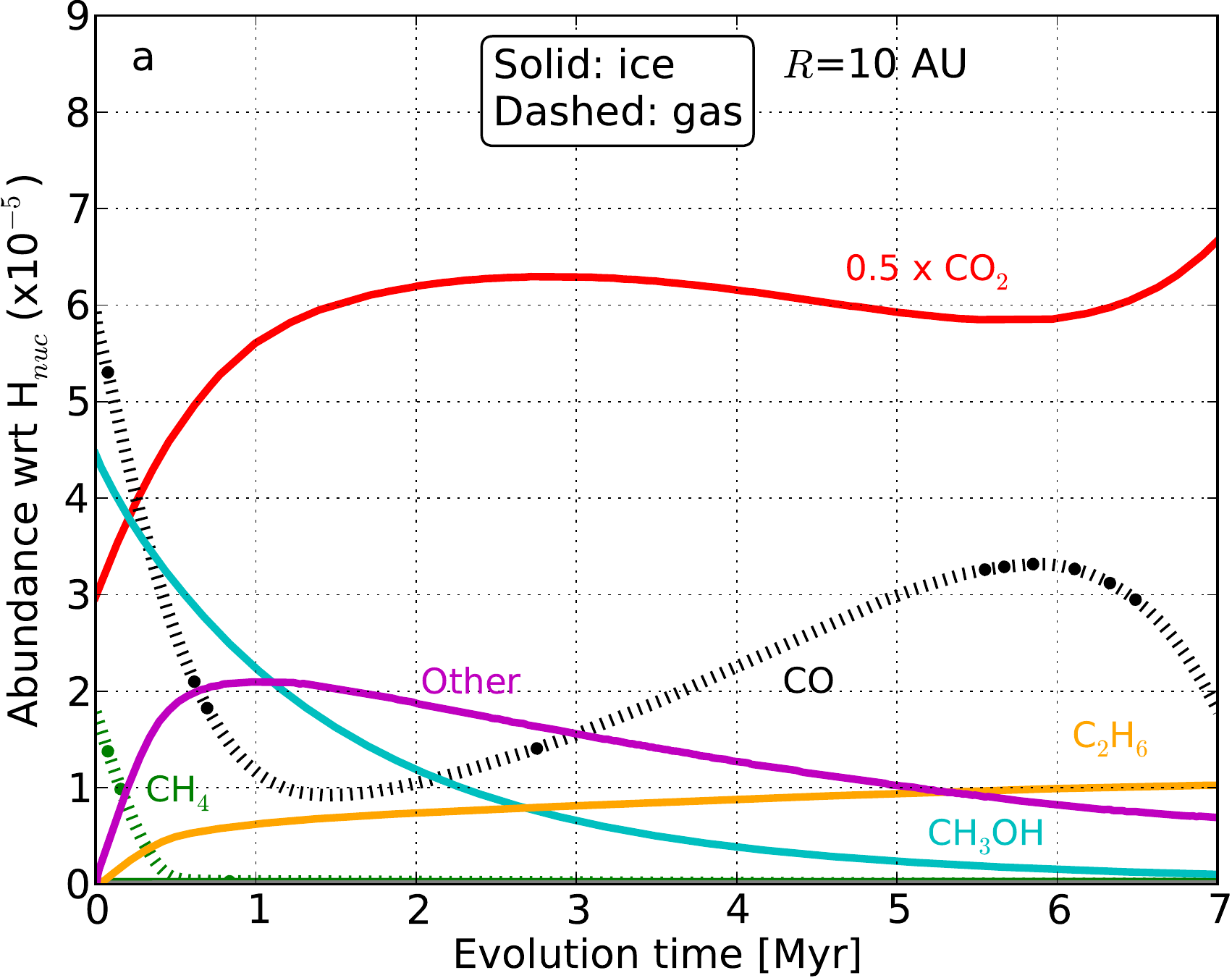}}
\subfigure{\includegraphics[width=0.3275\textwidth]{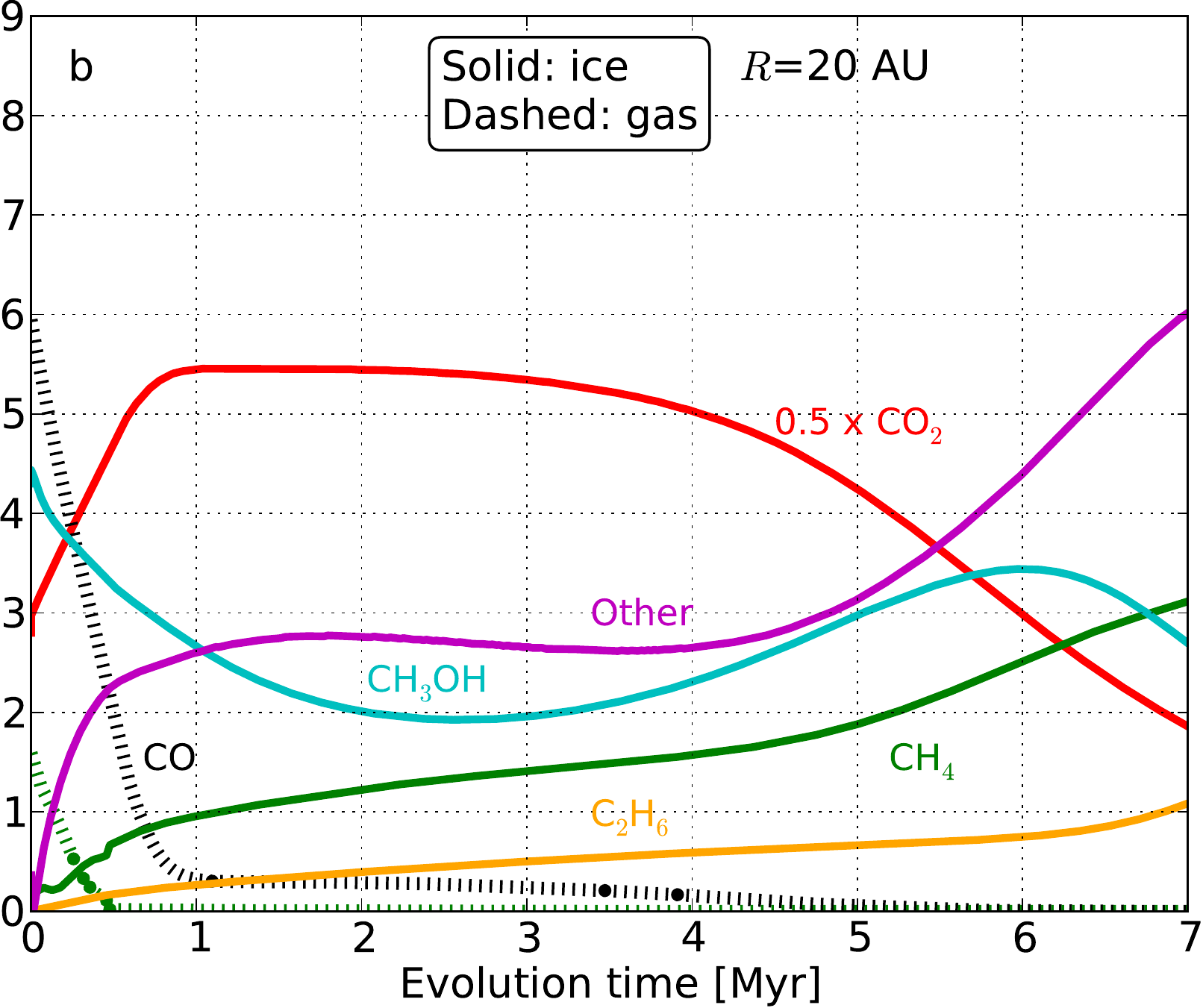}}
\subfigure{\includegraphics[width=0.3275\textwidth]{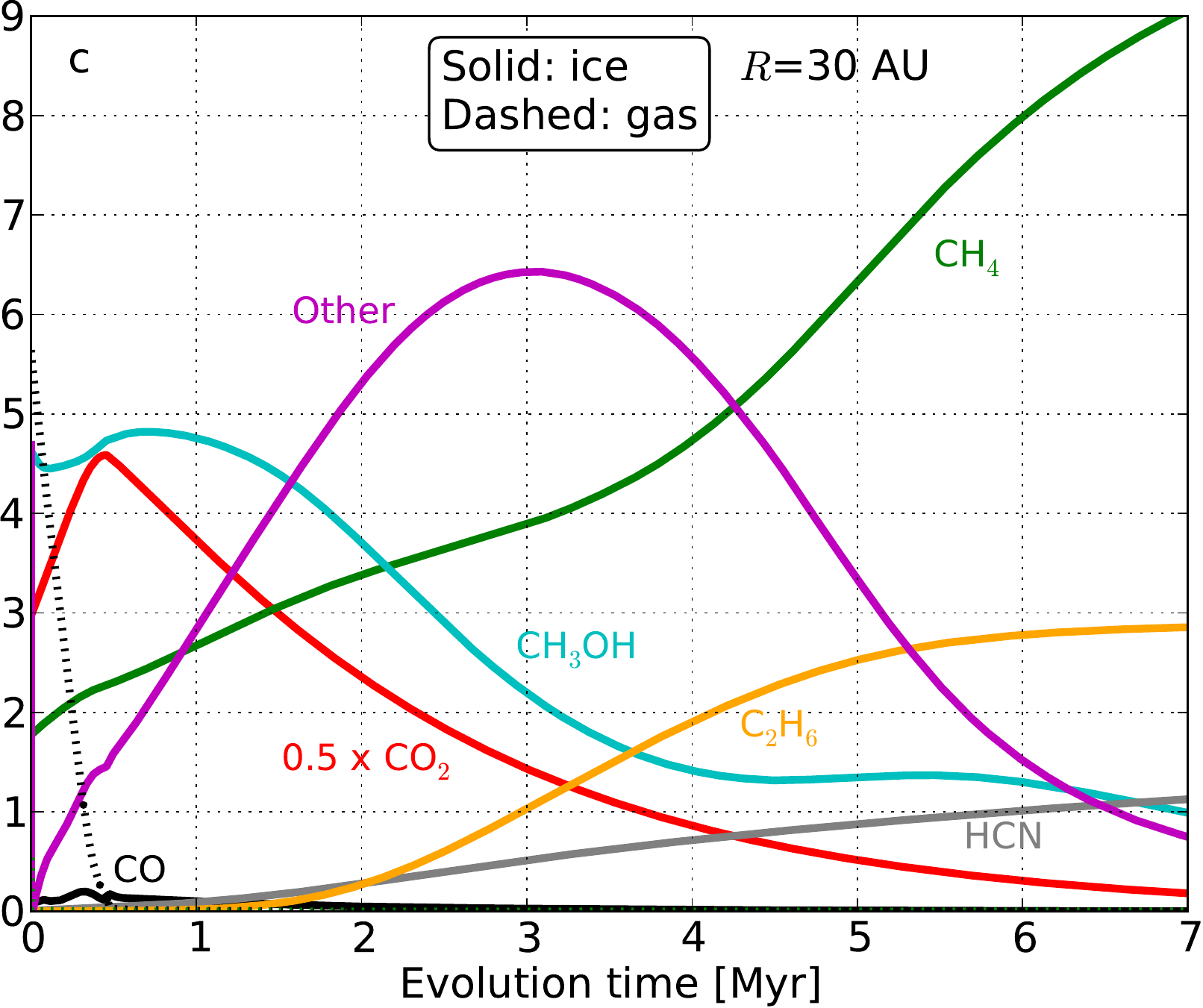}}
\caption{Time evolution of the main volatile species at 10, 20 and 30 AU, plotted on linear scales to ease comparison to Fig. 7 of \citet{yu2016} and Fig. 2 of \citet{yu2017}. The \ce{H2O} ice abundance in these plots is at least 80. Note the rapid decrease of gaseous CO within 1 Myr. Notice that \ce{CO2} ice is plotted at half its abundance to emphasise the steep drop for CO gas before 1 Myr at all three radii.}
\label{yu_comp}
\end{figure*}

In Fig. \ref{yu_comp} the abundance evolution of the main carbon-bearing volatiles, CO, \ce{CO2}, \ce{CH4}, HCN, \ce{C2H6}, \ce{CH3OH} are plotted at 10, 20 and 30 AU, alongside ``Other'', which accounts for the carbon that is not contained in the plotted species. For all three radii, the main contributors to ``Other'' are HCO and HNC, as well as complex molecules and hydrocarbons such as \ce{C3H4},  \ce{CH2CO}, \ce{H2CCO}, \ce{CH3OCH3} and \ce{CH3COCH3}, which all show abundances > 10$^{-7}$ with respect to H nuclei. It is evident at 10 AU in panel a that \ce{CO2} locks up the majority of the carbon, through the grain-surface reaction with CO and OH, and CO gas is seen to decrease in abundance. Here, \ce{C2H6} ice is accounting for about 10\% of the elemental carbon, with less than 10\% of the carbon locked up in ``Other''.

At 20 AU in panel b the picture looks different after 7 Myr. The ``Other'' species take up about 30\% of the elemental carbon, which is more than any single species alone. At 20 AU, the temperature decreases from about 30 to 18 K during the evolution time. That means that at the end of evolution, all molecules are frozen out onto the grains, and due to the temperature regime, most species are mobile on the grain surfaces, and able to react to form larger complex molecules, like those included in the ``Other'' category.

In panel c at 30 AU, the temperature goes down to 15 K after 7 Myr, meaning all species, including H atoms, can stick to the grains. H atoms, along with lowered mobility of molecules on the grain surfaces, means that hydrogenation reactions become dominant. Thus carbon is locked up in \ce{CH4}, HCN and \ce{C2H6} and \ce{CH3OH}, with only around 10\% elemental carbon in ``Other'' species. 

Recent work by \citet{yu2016,yu2017} also find this chemical conversion of CO into more complex ices using an evolving disk model in the midplane. Their disks have temperatures of $\sim$35-40 K at 20-30 AU. They also found CO destruction in the disk midplane, by modelling chemical evolution using the older {\sc Rate}06 network. However, their models produce the \ce{C2H5} radical in the ice, which in this work is hydrogenated into \ce{C2H6} ice. Also, they find only about 5\% increase in their \ce{CO2} ice abundance, indicating that the CO gas destruction is not feeding the production of \ce{CO2} ice on the grains. \citet{reboussin2015} did see conversion of CO gas to amongst other \ce{CO2} and \ce{C2H6} ices on the grains using the {\sc Nautilus} model, in agreement with this work. Besides these studies, models focusing on analysis of chemical evolution in disk midplanes are limited. \citet{cridland2017} used a chemical network to evolve chemistry in the midplane before forming planets out of the material, but did not focus on the chemical response to the (varying) physical conditions and reaction types dominating in the midplane. \citet{helling2014} found a different effect of chemistry on the C/O ratios that what is seen in this work, because their treatment of grain-surface chemistry was simpler.

\subsection{Evolving elemental ratios}
\label{disc_co}

\begin{figure*}
\subfigure{\includegraphics[width=0.5\textwidth]{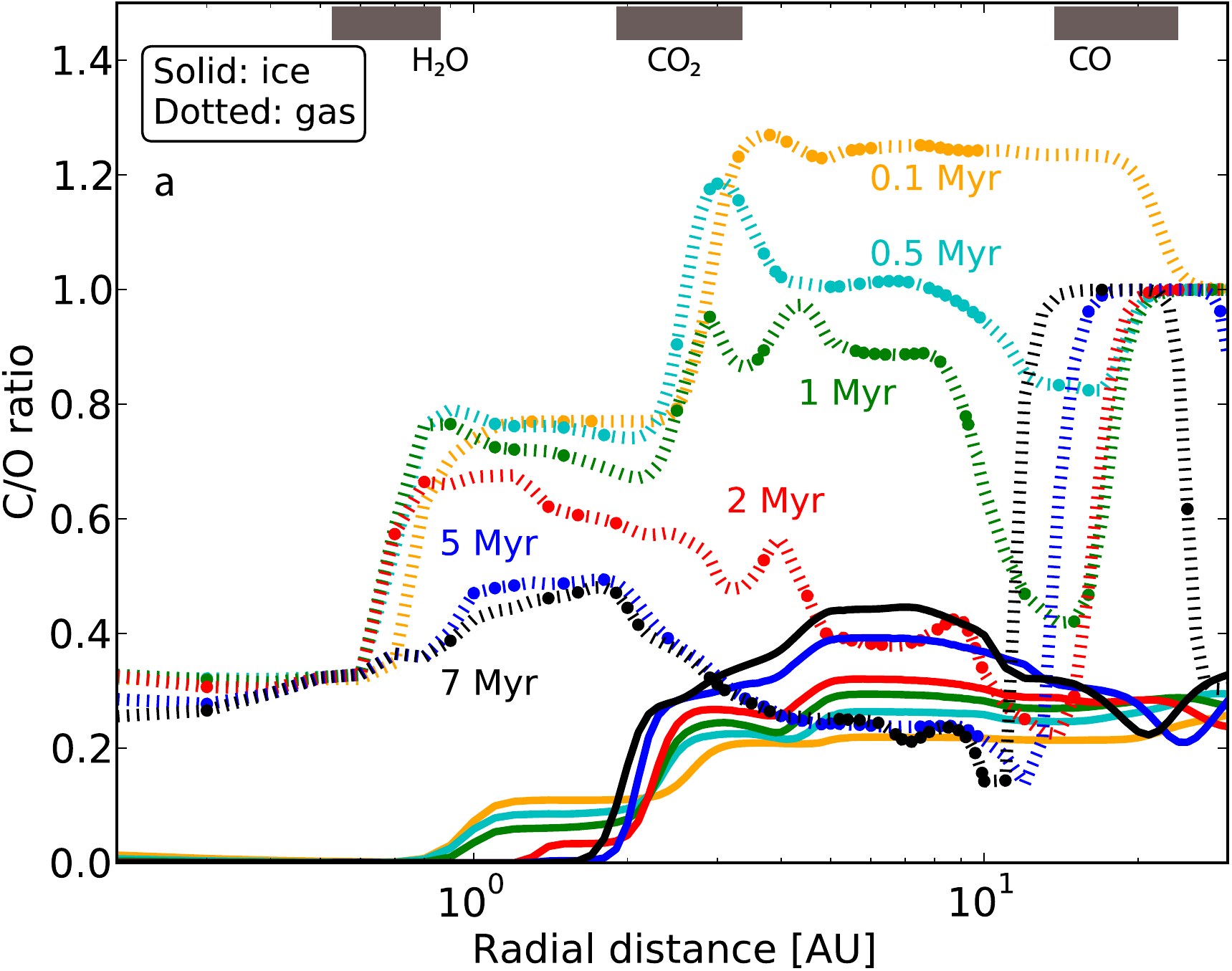}}	
\subfigure{\includegraphics[width=0.5\textwidth]{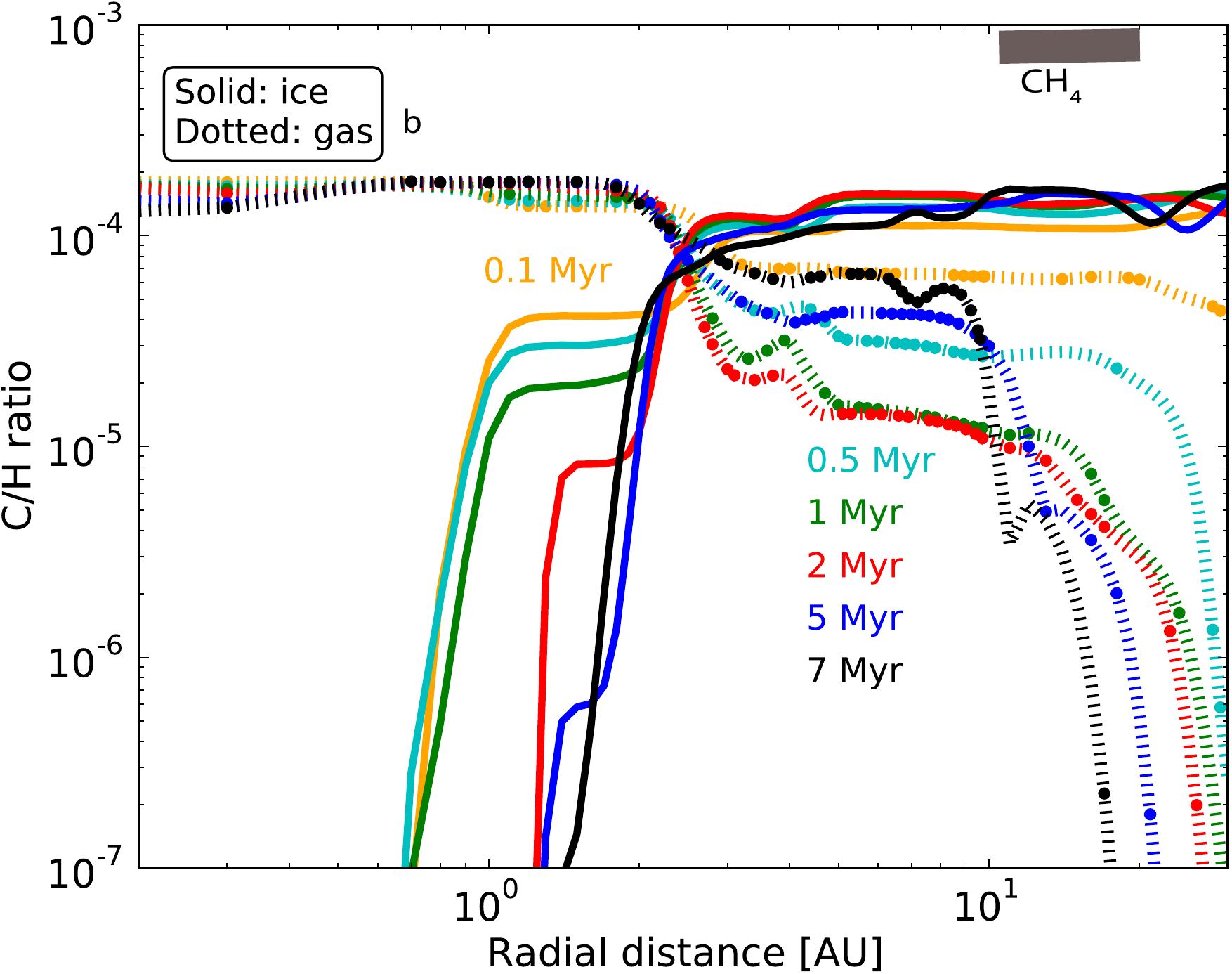}}\\
\subfigure{\includegraphics[width=0.5\textwidth]{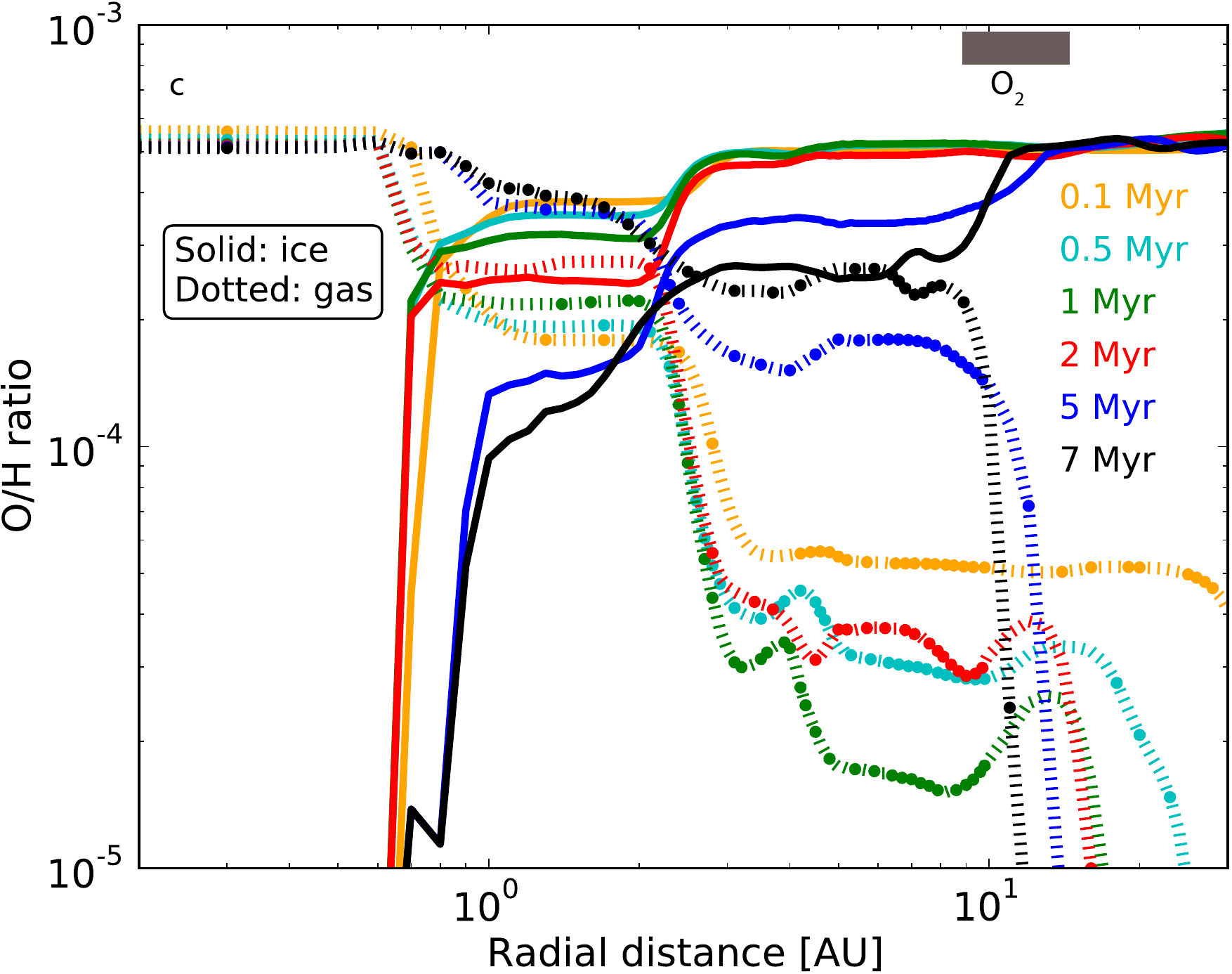}}
\subfigure{\includegraphics[width=0.5\textwidth]{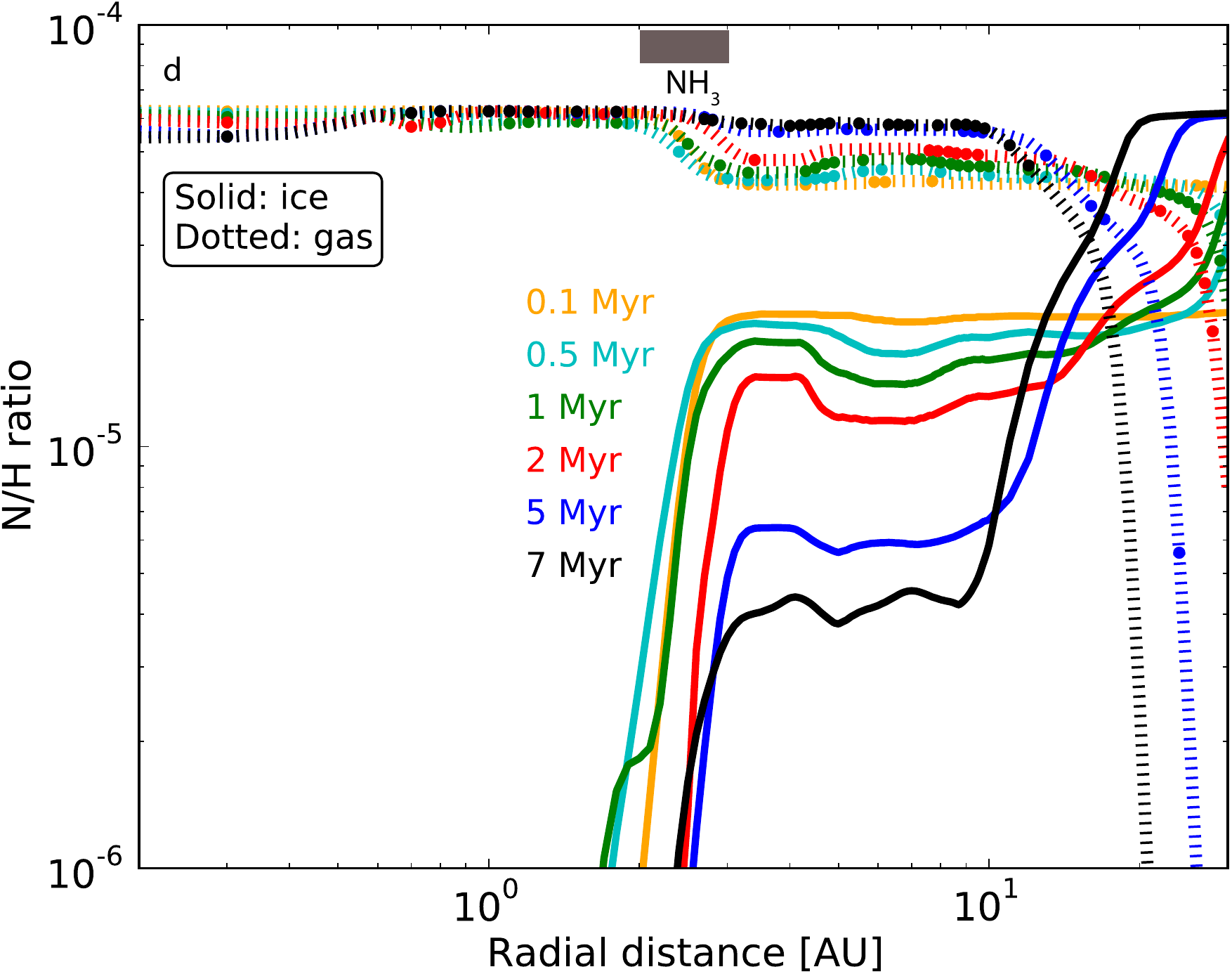}}\\
\caption{Elemental ratios of carbon, oxygen and nitrogen in gas and ice as a function of radius, at six timesteps: 0.1, 0.5, 1, 2, 5 and 7 Myr. Panel a: C/O ratios. Panel b: C/H ratios. Panel c: O/H ratios. Panel d: N/H ratios. Gas profiles are dashed, ice profiles are solid. Note the different scales on the $y$-axes, in particular the linear scale in panel 1. The grey boxes indicate the radial regions that the iceline of a volatile move over during the evolution.}
\label{elem_ratios}
\end{figure*}

An important aspect of chemical evolution in a disk midplane is the relation to the composition of exoplanets, particularly their atmospheres. In Paper 1 it was shown that unless the inheritance scenario with low ionisation was considered, the C/O ratio of the chosen model setup changed over the course of 1 Myr of evolution. It was demonstrated that the C/O ratio radial ``stepfunction'' in Fig. 1 of \citet{oberg2011co} was modified.  In this work, the evolution of the C/O ratios are traced up to 7 Myr instead of 1 Myr, and the effects of both chemical evolution and the evolving temperature structure are shown.

The longer timescale causes larger changes to the dominant volatile abundances and significant changes to the traditional C/O ratio ``stepfunction'', even more than what was seen Paper 1. The larger abundance changes are apparent in Fig. \ref{evolutions_lowmass}, where changes of up to 2 orders of magnitude are seen from 1-7 Myr, thus significantly larger than the changes happening until 1 Myr (the case from Fig. 4 in Paper 1).
Considering the more realistic scenario of an evolving, cooling disk (see Fig. \ref{inh_vs_reset} panels b and d) rather than a static one (Fig. \ref{inh_vs_reset} panel c), similar trends in chemical abundance changes to the static case are seen. The main effect is that volatile icelines are shifted inwards with time, because the disks cool. The physical conditions at each timestep of the simulations determine the chemical composition at that timestep. This holds for all scenarios and tells us that the chemical timescale is faster than the physical timescale in these models, and that a static disk model may be sufficient for modelling chemistry. However, a planet forming in the crossing-zone of an iceline may experience a drastic change to the composition of accreting material over time.

In Fig. \ref{elem_ratios}a, b, c and d, the evolving C/O, C/H, and O/H and N/H ratios, respectively, in gas and ice are presented for the evolving 0.1 MMSN disk. The yellow profiles are at 0.1 Myr only, where as yet no significant chemical evolution has taken place, and the ratios are therefore the canonical values corresponding to the input abundances. Thus, for C/O for example, this yellow profile resembles the ``stepfunction'' from \citet{oberg2011co}. 

For the C/O ratio evolution in panel a, it is clear at later times that the evolving chemistry has caused the C/O ratios in gas and ice to change. At 5 AU, the C/O ratio in the gas drops from 1.2 to 0.3 over the course of 7 Myr. For the ice at the same distance, the C/O ratio increases from 0.2 to almost 0.5, thus making the ice between 3 and 10 AU more carbon rich than the gas. By 7 Myr the C/O ratios in both gas and ice are below 0.5, and thereby the C/O gas and ice ratio profiles are significantly different than the \citet{oberg2011co} stepfunction, and the profiles correlate with the C/O ratio profiles for the reset scenario from Paper 1. That indicates that steady state is close after 7 Myr of evolution with inherited abundances. It is expected that so long as the initial C/O<1, the same trends would be seen in C/O ratio evolutions, if a different set of elemental ratios were chosen.

The C/H evolution in panel b shows a significant turnover at the \ce{CO2} iceline at 2 AU, with more carbon being contained in gas than in ice inside the iceline, and vice versa outside of it. This iceline is thus critical for the metallicity of the material, and has important implications for the carbon content of the forming planets. Besides this distinct iceline, other features of the C/H ratios are the decreasing level in ice between the \ce{CH3OH} (at 1.3 AU by 7 Myr) and the \ce{CO2} icelines. This is due to the destruction of \ce{CH3OH} on the grain surfaces, as well as the CO and \ce{CO2} dependent interconversion of C/H ratio in gas and ice between the \ce{CO2} and the \ce{O2} icelines (see Section \ref{disc_timescales}).

Panel c features the O/H ratio evolution. Here the \ce{H2O}, \ce{CO2} and \ce{O2} icelines at $\sim$0.8, 2 and 10 AU, outline the radii of main changes. Between the \ce{H2O} and \ce{CO2} icelines, the ice starts as two times more oxygen rich than the gas, but after 7 Myr, the gas can become 3 times more oxygen rich than the ice. Between the \ce{CO2} and \ce{O2} icelines, the ice starts an order of magnitude more oxygen rich than the gas, but after 7 Myr, the gas and ice are equally oxygen rich. This is due to the increasing abundance of gaseous \ce{O2}. 

Although this work so far has focused on carbon and oxygen-bearing species, for completeness the evolving nitrogen content in the material is shown in panel d. The decreasing nitrogen ice content between 2 and 8 AU is due to the destruction of \ce{NH3} ice (iceline at $\sim$ 2.5AU) in this region, which causes an increased nitrogen level in the gas, with \ce{N2} (iceline at $\sim$20 AU) becoming the main nitrogen gas carrier (see Paper 1 for further discussion). \citet{schwarz2014} also modelled the evolution of nitrogen-bearing species in a disk, and found different partitionings of the nitrogen into HCN, \ce{NH3} and \ce{N2}, depending on their model assumptions.


\subsection{Impact for planet formation and planet atmospheric composition}
\label{disc_planets}

Recent years has seen enormous effort put into the field of planet formation and planet population synthesis models. Physical effects like grain growth and migration/radial drift, and a planet's acquisition of an atmosphere from a disk have been included in increasingly sophisticated models \citep[see e.g.][and references therein]{lambrechts2012,alibert13,alidib2014,madhu2014,benz2014,bitsch2016}. Planet population synthesis models however have neglected chemical evolution, although \citet{cridland2016,cridland2017} did consider kinetic chemistry to model the material that go into forming planetary atmospheres.

 \citet{mordasini2016} conducted planet formation synthesis models introducing a "chain" of models to self-consistently predict the atmospheric transmission spectra of a hot Jupiter exoplanet from modelling the formation of the planet as it forms in the disk. One prediction from their work was that for hot Jupiters less massive than a few $M_{\rm{Jup}}$, heavy element enrichment in the envelope from impacting planetesimal polluters dominate the final atmospheric composition over the enrichment obtained from gas accretion. This is interesting in the context of the findings in this work, that the C/O, C/H, O/H and N/H ratios in both gas and ice evolve over time. For hot Jupiters, it is thus not necessarily the heavy element content of the gas in the disk that will determine the planet's resulting atmospheric elemental ratios. 
 
The findings of this work, that chemistry matters in disk midplanes under certain assumptions, may complicate the challenge of relating observed exoplanet atmospheric elemental ratios to the formation history of the planet. Not only is it shown in this paper that chemical evolution can change the radial elemental ratios in gas and ice over time, but as \citet{mordasini2016} stated, the disk material - gas or ice - that the atmosphere of the planet is formed from, may have a mixed origin.

The result of these two effects can, however, be explored through modelling. By having a planet atmosphere form from different relative contributions of gaseous and icy material in the disk, and by assuming (as a start) one of the time evolved elemental ratio abundances presented here, the final elemental ratio in the atmosphere, and the resulting transmission spectrum, can be generated. It can then be assessed whether or not the effects are observable, with current or future observational facilities. This would also be a first step to better implement disk chemistry into the ``chain'' of models, as requested by \citet{mordasini2016}. First attempts at this were indeed carried out by \citet{cridland2016,cridland2017}. However, they only considered planets formed inside 4 AU. An expansion of their work would thus be welcomed.

Lastly, the outer disk volatile ice abundances presented here obviously inspire a comparison to observed abundances in comets in the Solar System, in order to constrain the models with Solar System conditions. Such a comparison is currently in progress and will be published in an upcoming paper.

\section{Summary}
\label{summary}

In this paper, the implications of chemistry in evolving disk midplanes on the 
composition of planet-forming material have been investigated. The main conclusions are:
\be
\item Chemical evolution is an ionisation-driven long-term effect, which for the 
disk structures considered here becomes significant 
after a few times 10$^{5}$ yrs of evolution, and goes on until at least 7 Myr, in the region between the \ce{H2O} and \ce{O2} icelines ($\sim$0.5-10 AU, for the 0.1 MMSN disk).
\item Considering the more realistic scenario of an evolving, 
cooling disk rather than a static one, similar trends in chemical abundance changes to the static case are seen. The main effect is that volatile icelines are shifted inwards with time, because the disks cool. Hence, using a static disk structure representative of the conditions in the disk during the accretion epoch of a forming planet may be sufficient for modelling chemistry.
\item Changes of similar magnitudes are seen for all scenarios for the 0.1 MMSN disk and the 0.55 MMSN disk. That means that assuming different disk masses does not significantly affect the chemical evolution. However, higher mass disks than those studied here should be explored.
\item After 7 Myr of evolution, inside the \ce{H2O} iceline and outside the \ce{O2} iceline, 
the fingerprint of interstellar abundances has been erased, for the high ionisation level, and chemical steady state has been achieved. That is to say, for ice in the outer disk, it becomes irrelevant 
whether the inheritance or the reset scenario is assumed. 
In the inner tenth of an AU, chemical equilibrium conditions are likely achieved. 
\item An abundance ``inversion'' is seen for CO and \ce{CO2} between the \ce{CO2} and CO icelines, where the abundance evolutions invert at a time which depends on the ionisation level. For the high ionisation level, CO gas reaches its minimum abundance by $\sim$1.6 Myr, and \ce{CO2} reaches its maximum abundance at $\sim$2.5 Myr.
\item By 7 Myr CO gas is seen to deplete inside the CO thermal iceline, from outside the \ce{O2} iceline, which could mimic the iceline being further in and at warmer temperatures than the CO binding energy prescribes.
\item Ice species such as HCN and \ce{C2H6} become important carbon-bearing 
species in the outer disk by a few Myr. Between the \ce{H2O} and \ce{O2} icelines, \ce{O2} gas 
is the dominant oxygen bearing species. Just outside the \ce{O2} iceline, it is warm enough for molecules to move on the grain surfaces, but not cold enough for H atoms stick on the grains. This results in the production of larger, more complex molecules on the grain surfaces.
\item Chemical evolution changes the elemental ratios in the gas and ice significantly over time. This results in a region where the ice has a higher C/O ratio than the gas. In order to relate observed atmospheric 
C/O ratios in exoplanets to formation locations and conditions in the 
disk midplane, chemical evolution is important to consider, 
independent on whether an atmosphere is built primarily from gas 
accretion or from planetesimal impacts. Generally, chemical evolution acts to significantly lower the C/O ratio of gas, such that neither the ratio in gas or in ice is at or above unity, but rather the C/O ratios in both gas and ice evolve to be below 0.5 (given an overall C/O ratio of 0.34).
\ee

If planet formation lasts for longer than a few times 10$^{5}$ yrs in the disk midplane and cosmic ray ionisation is present, then the chemical abundances and elemental ratios in gas and ice will change alongside with planet formation. Henceforth, chemical evolution needs to be addressed when predicting the chemical makeup of planets and their atmospheres, as well as when inferring formation locations of exoplanets based on their observed atmospheric elemental ratios.

\begin{acknowledgements}

The authors thank Amaury Thiabaud and Ulysses Marboeuf for making their disk density and temperature profile available and for many useful discussions. The authors also thank Nikku Madhusudhan for a helpful discussion regarding the C/O ratio evolution and its effect on the metallicity, and the referee for useful comments that improved the presentation of the paper. Astrochemistry in Leiden is supported by the European Union A-ERC grant 291141 CHEMPLAN,
by the Netherlands Research School for Astronomy (NOVA), and by a Royal Netherlands Academy of Arts and Sciences (KNAW) professor prize. CW also acknowledges the Netherlands Organisation for Scientific Research (NWO, grant 639.041.335) and the University of Leeds for financial support.

\end{acknowledgements}
\bibliographystyle{aa} 
\bibliography{bib_new} 
\begin{appendix}
\label{app_high_mass}
\section{Physical structure for 0.55 MMSN disk}
\begin{figure*}
\subfigure{\includegraphics[width=0.5\textwidth]{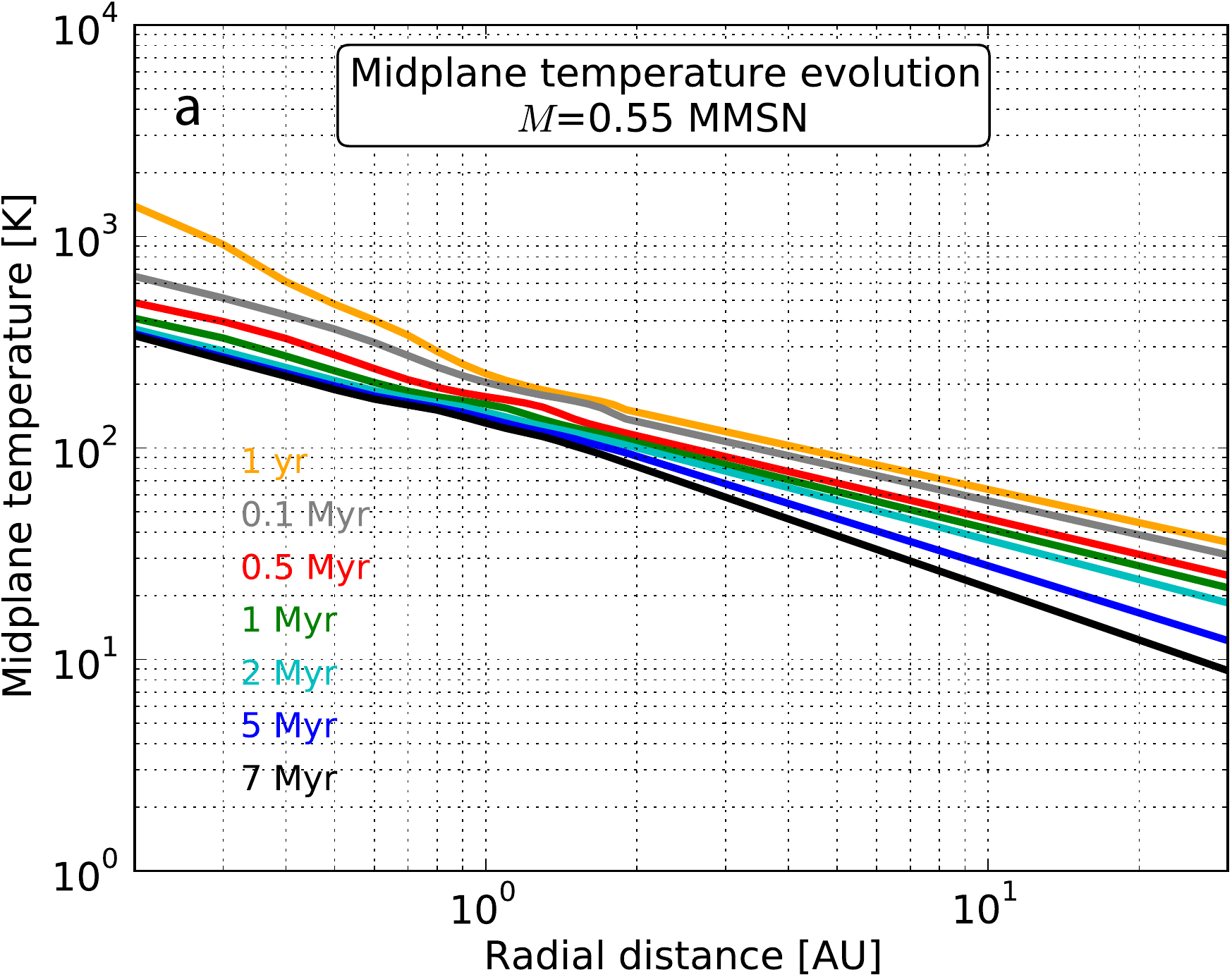}}
\subfigure{\includegraphics[width=0.5\textwidth]{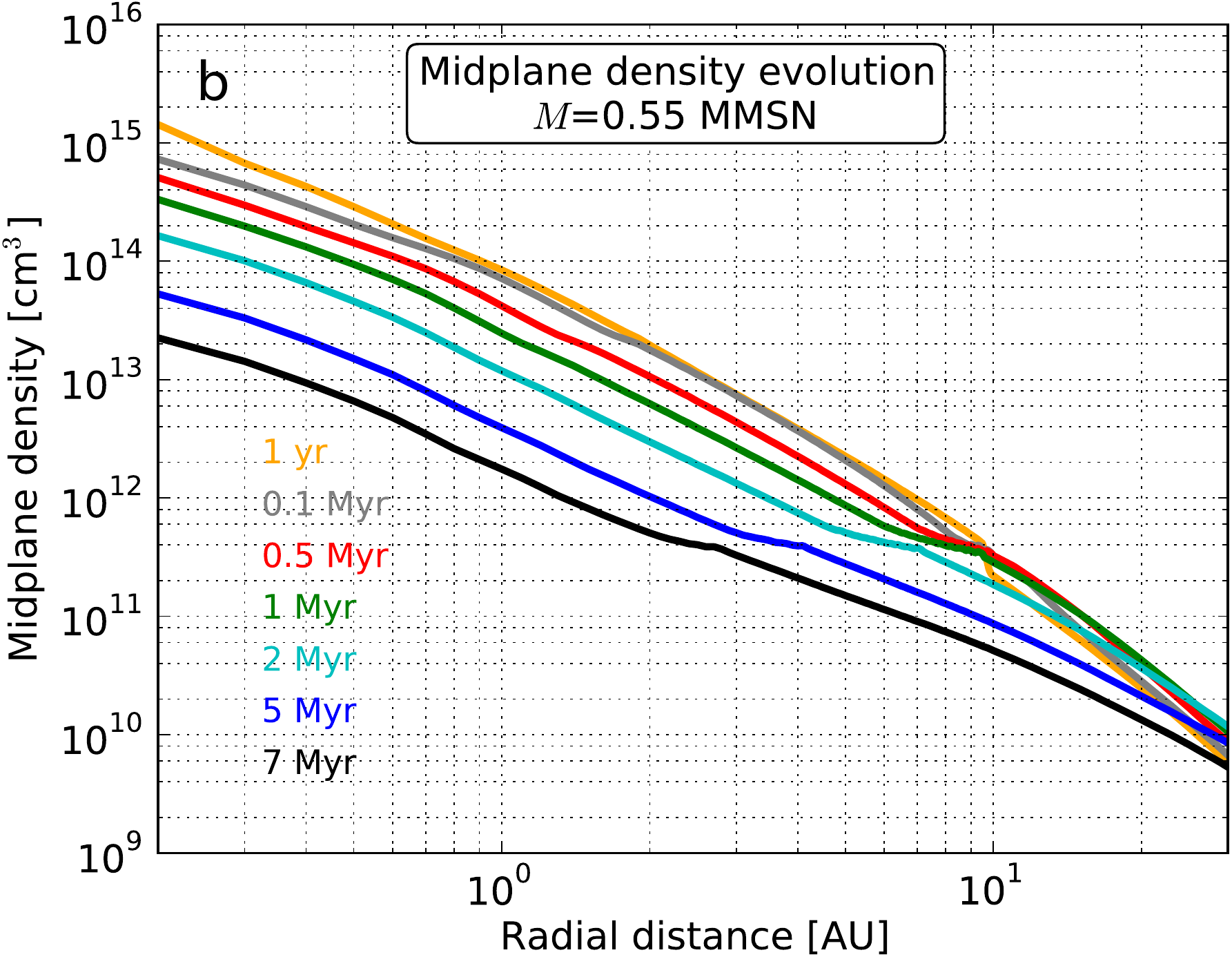}}\\
\subfigure{\includegraphics[width=0.5\textwidth]{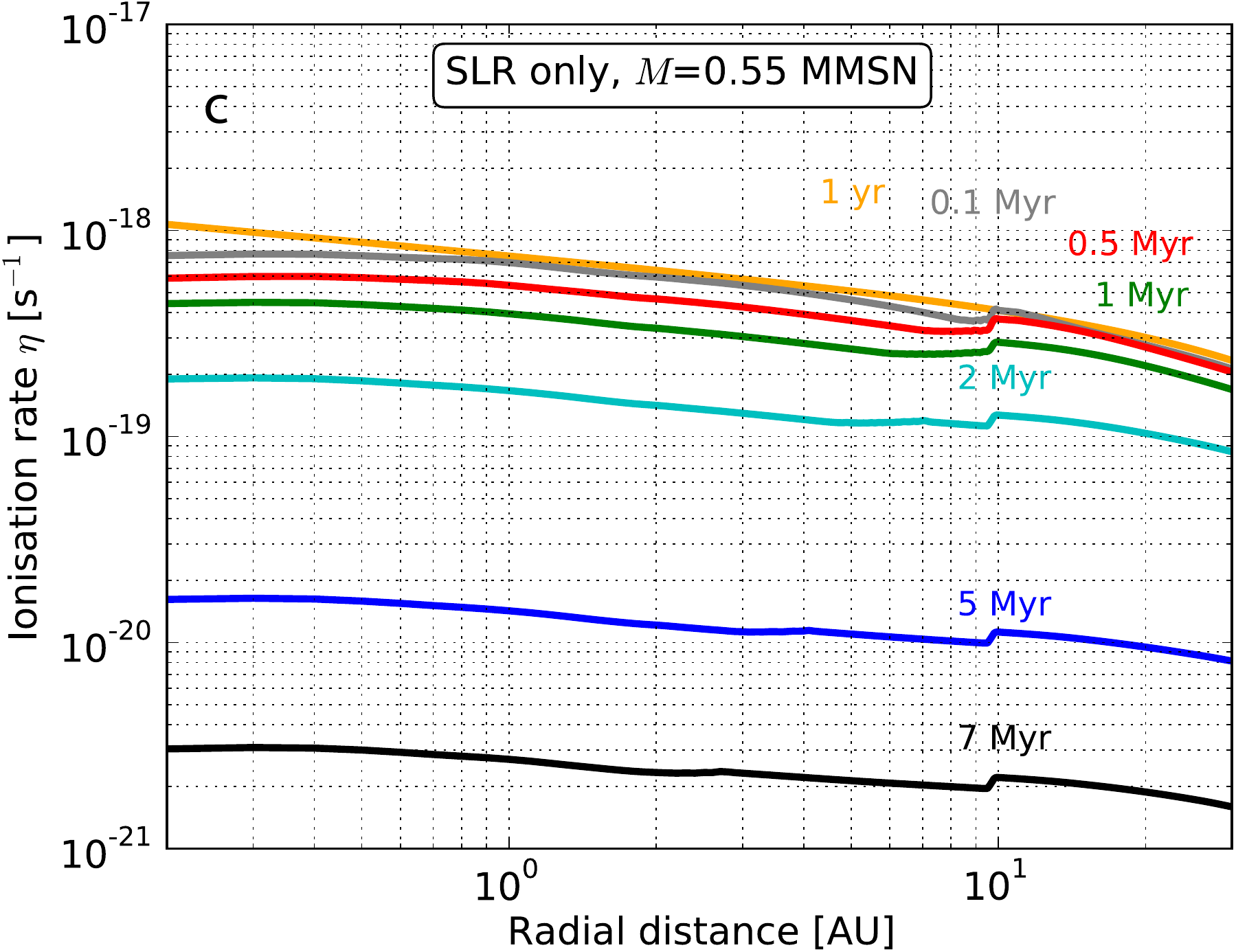}}
\subfigure{\includegraphics[width=0.5\textwidth]{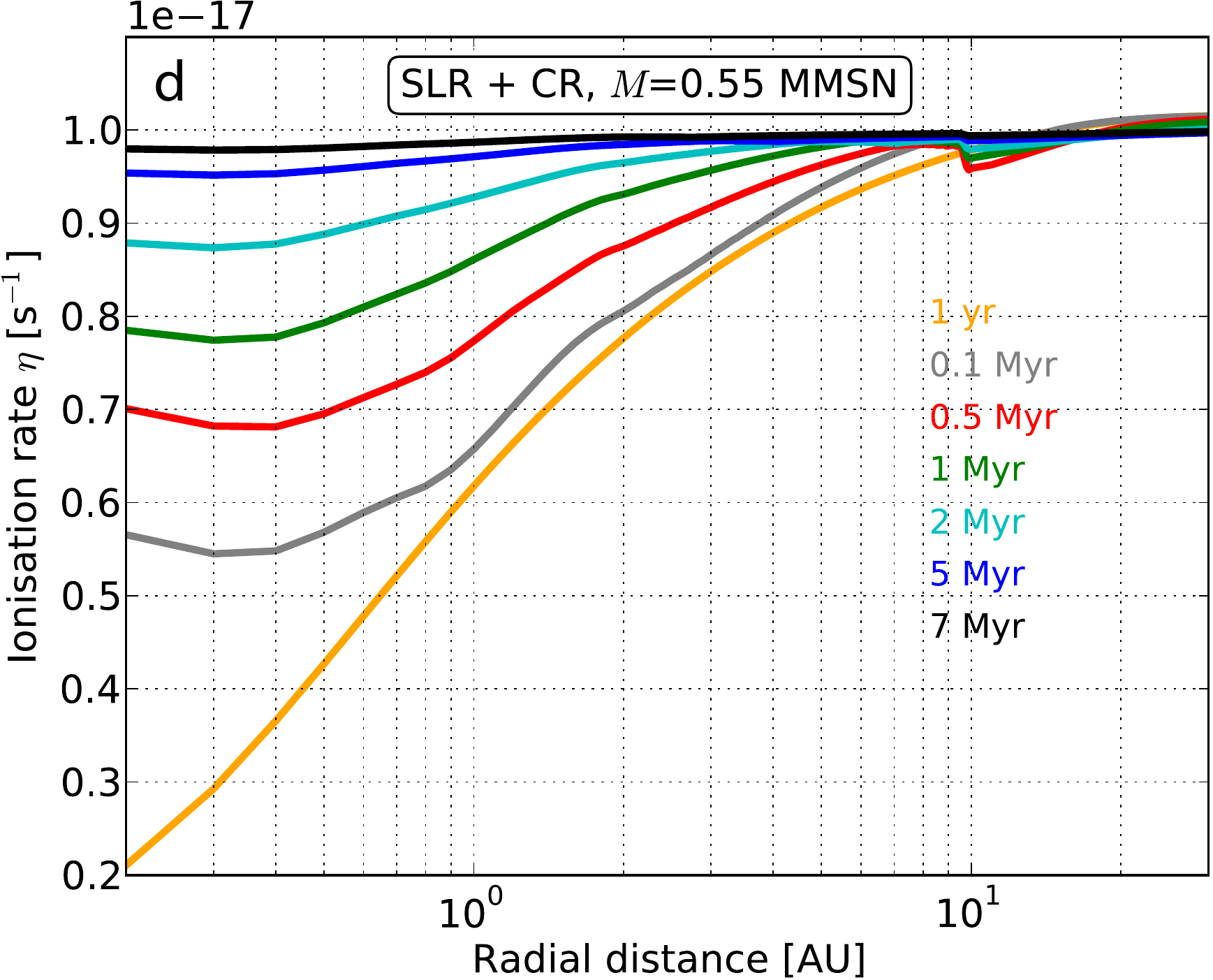}}\\
\caption{Same as for Fig. \ref{low_mass_phys}, but for the 0.55 MMSN disk.}
\label{high_mass_phys}
\end{figure*}

Figure \ref{high_mass_phys} show the evolving physical structure for the 0.55 MMSN disk, featuring temperature, density and ionisation level evolution.

\end{appendix}
\end{document}